\documentclass[12pt]{article}
\pdfoutput=1
\usepackage[hmargin={2.5cm, 2.5cm}, vmargin={2.5cm, 2.5cm}]{geometry}
\usepackage{caption}
\usepackage{subcaption}
\usepackage{hyperref}
\usepackage{url}
\usepackage{graphicx}
\usepackage{float}
\usepackage{verbatim}
\usepackage[affil-it]{authblk}

\providecommand{\keywords}{\small\textbf{Keywords: }}
\DeclareUnicodeCharacter{2212}{\textendash}

\newenvironment{acknowledgements}{%
  
  \begin{abstract}
}{%
  \end{abstract}
}

\title{\large \textbf{Probing Gluons at the Spin Physics Detector}}

\author[1,a)]{\small A. Guskov}
\author[1,b)]{A. Datta}
\author[2,c)]{A. Karpishkov}
\author[1,d)]{I. Denisenko}
\author[2,e)]{V. Saleev}

\affil[1]{\footnotesize Joint Institute for Nuclear Research, Joliot-Curie 6, Dubna-141980, Moscow Region, Russia.}
\affil[2]{Samara National Research University, Moskovskoye Hwy 34, Samara-443086, Samara Region, Russia}
\affil[a)]{avg@jinr.ru}
\affil[b)]{amareshdatta@gmail.com}
\affil[c)]{karpishkoff@gmail.com}
\affil[d)]{iden@jinr.ru}
\affil[e)]{saleev@samsa.ru}

\date{}

\begin{document}
\maketitle

\begin{abstract}
 The Spin Physics Detector (SPD) at the Nuclotron based Ion Collider fAcility (NICA) is a multi-purpose experiment designed to study nucleon spin structure in the three dimensions. With capabilities to collide polarized protons and deuterons with center of mass energy up to 27 GeV and luminosity up to $10^{32} \rm cm^{-2} \ s^{-1}$ for protons (an order of magnitude less for deuterons), the experiment will allow measurements of cross-sections and spin asymmetries of hadronic processes sensitive to the unpolarized and various polarized (helicity, Sivers, Boer-Mulders) gluon distributions inside the nucleons. Results from the SPD will be complimentary to the present high energy spin experiments at the RHIC facility or future experiments like the EIC (at BNL) and the AFTER (at LHC) in understanding the spin structure of the basic building blocks of visible matter. Monte Carlo simulation based results presented here demonstrate the impact of the SPD asymmetry mesurements on gluon helicity PDF and gluon Sivers functions. With polarized deuteron collisions, the SPD will be the unique laboratory for probing tensor polarized gluon distributions. In addition, there are also possibilities of colliding other light nuclei like Carbon at reduced collision energy and luminosity at the first stage of the experiment.
\end{abstract}


\section{Introduction}

Over the last few decades, experimental results have often surprised the physics community and opened up new windows to the intricate details of the structure of the fundamental building blocks of Nature. European Muon Collaboration (EMC) results \cite{ref-EMC} shed light on the importance of the possible gluonic contributions to the nucleon spin. E704 and other results \cite{ref-E704, ref-large_AN} of large single spin asymmetries inspired the community to think of the motions of the quarks and gluons inside the nucleons. 

Visible matter made of quarks and gluons is mostly described with the help of Quantum Chromodynamics (QCD), the theory of strong force. Our present understanding of the quarks and gluons comes from the high energy limit of perturbative QCD (pQCD) \cite{ref-QFact, ref-GFact}. Decades of experimental measurements of inclusive and semi-inclusive Deep Inelastic Scattering (DIS) (at COMPASS, HERMES), electron positron scattering (at HERA), hadron scattering (at RHIC) have so far given us a fairly precise description of quarks \cite{ref-NNPDF, ref-NNPDF4} inside the nucleons using pQCD as the preferred tool for interpretations. However, the gluonic component (which accounts for $\sim 99\%$ of all nucleon mass and therefore, all visible matter) is still poorly understood. It is imperative for the physics community to experimentally access the gluons inside the nucleons to be able to consistently describe the baryonic matter and their interactions.

Gluon distributions inside nucleons are harder to access than those of the quarks in semi-inclusive DIS scattering of leptons off hadrons as gluons do not interact with leptons directly via strong force. Hadronic scattering at high energies has been, in the recent years, the best tool for probing gluon spin distributions inside protons \cite{ref-DSSV2014}. Understanding of the gluon helicity distributions have changed over the first couple of decades of the twenty first century as the analyses included more and more experimental data form various sources \cite{ref-GRSV, ref-BB, ref-MSTW, ref-DSSV}.

A more complete picture of the three dimensional partonic structure has been emerging \cite{ref-TMD-overview} in the last decade or so with more and more data to access the transverse momentum dependent (TMD) parton distribution functions (PDF). Large transverse asymmetries in hadron production necessitated a closer look at the partonic structure including transverse momentum dependent distribution functions \cite{ref-Sivers} and fragmentation \cite{ref-Collins}.

The SPD \cite{ref-CDR} is a proposed experiment at the Nuclotron based Collider fAcility (NICA) at the Joint Institute for Nuclear Research (JINR) in Dubna. It is particularly focused at probing the gluons inside protons and deuterons. The SPD will make cross-section and asymmetry measurements of several hadronic processes sensitive to various (unpolarized and polarized) gluon distributions.

\begin{figure}[H]
\begin{subfigure}[h]{0.5\textwidth}
 \includegraphics[width=\textwidth]{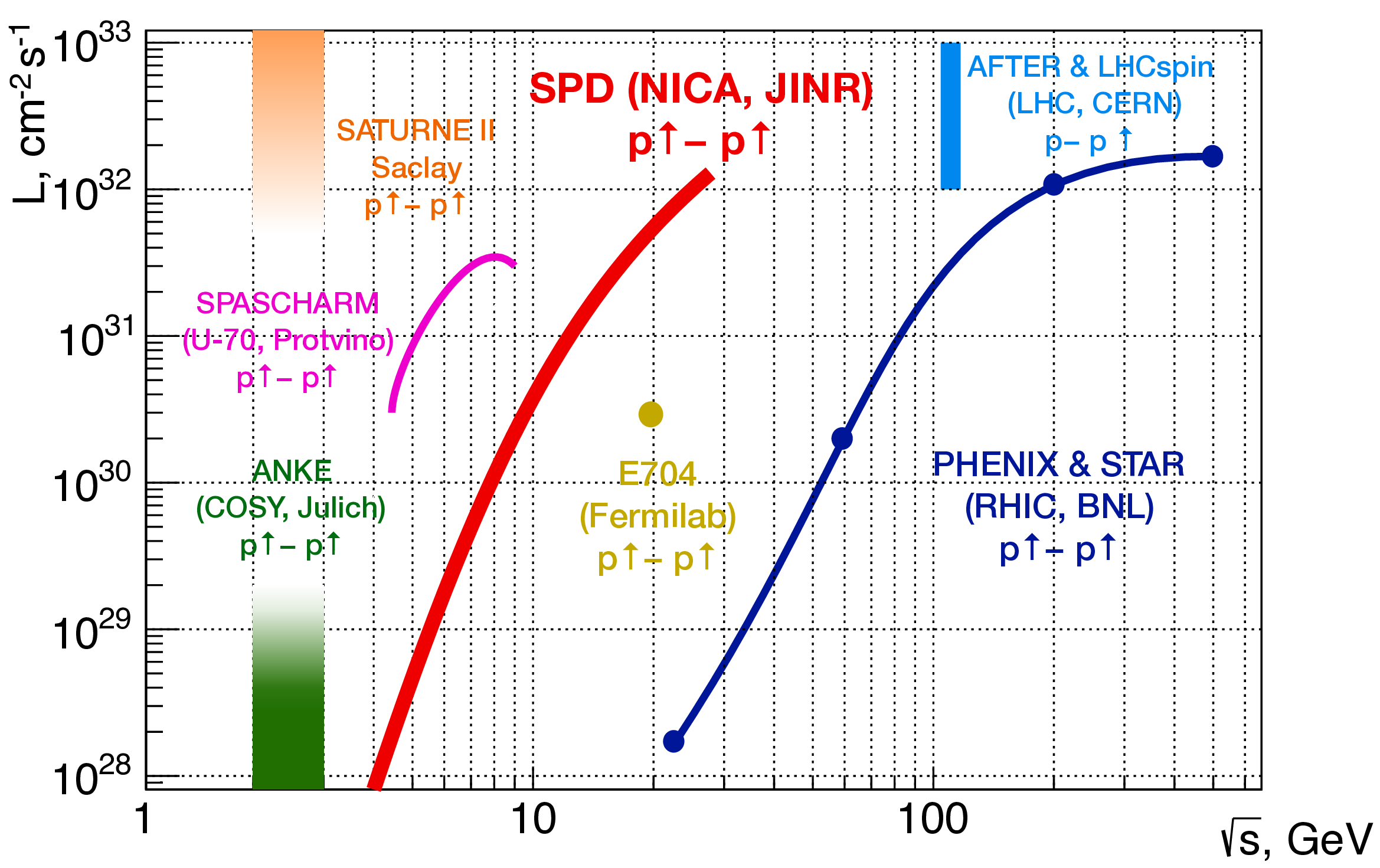}
 \caption{\centering  }
 \label{fig0a}
\end{subfigure}
\hfill
\begin{subfigure}[h]{0.5\textwidth}
\centering
 \includegraphics[width=0.85\textwidth]{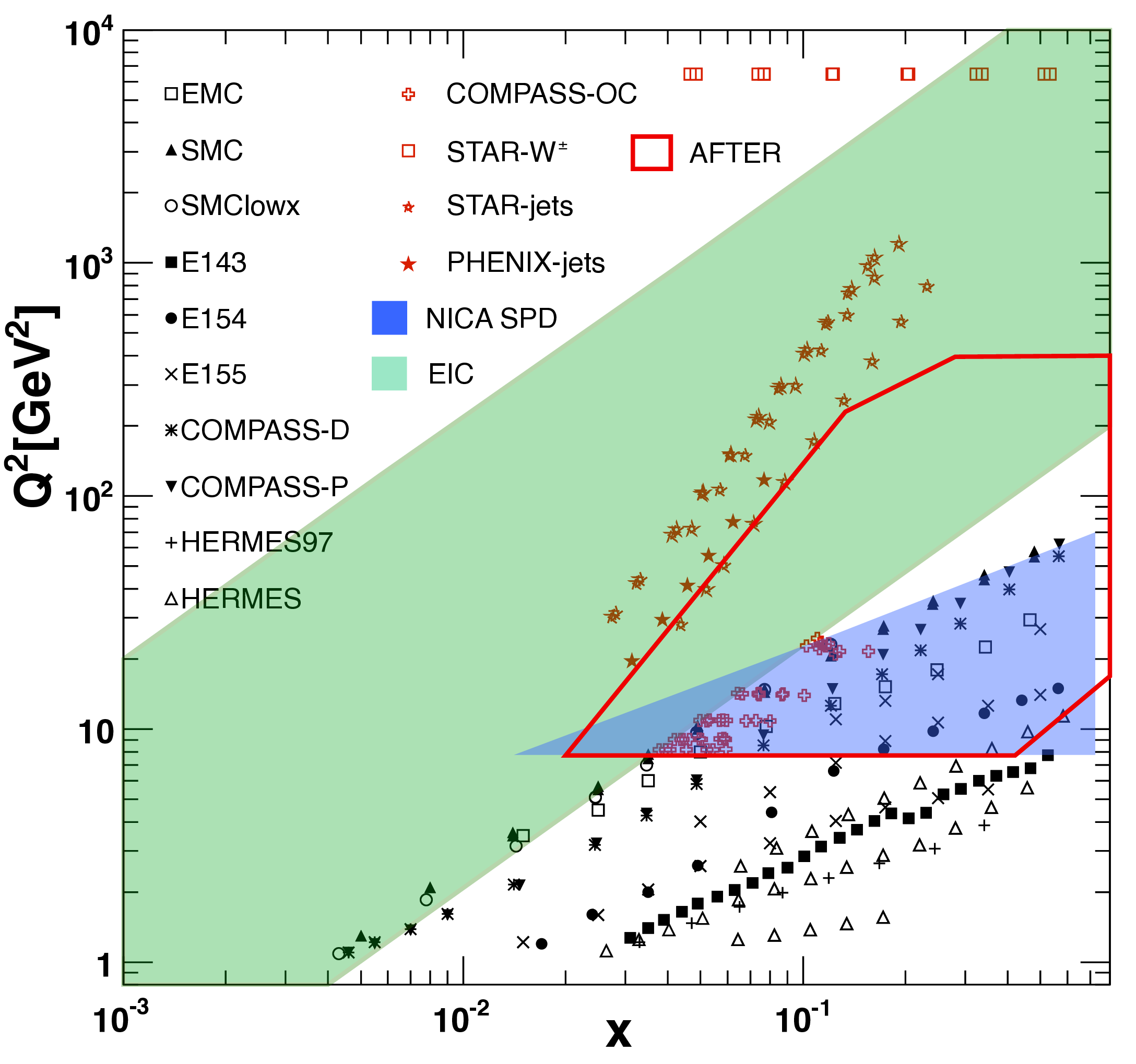}
 \caption{\centering  }
 \label{fig0b}
\end{subfigure}
\caption{(a) Luminosity and centre of mass energy of collision for the SPD and other relevant spin experiments. (b) Kinematic coverage of the SPD and other future spin experiments.}
\label{fig0}
\end{figure}

The SPD will operate at medium energy ranges (up to 10GeV in the initial stage  and up to 27 GeV in the later stage) that are complementary to the present and future experiments (Figure \ref{fig0a}) with higher center-of-mass energies (i.e. PHENIX, STAR, AFTER \cite{ref-LHC-AFTER}, LHCspin \cite{ref-LHCspin}). As a consequence, measurements at the SPD will probe high momentum fraction $x$ and low to medium energy scale $Q^2$ that will provide access to gluonic distribution in a kinematic regime (illustrated by Figure \ref{fig0b}) that is complementary to those accessed in other upcoming major spin physics experiment like the Electron Ion Collider (EIC) \cite{ref-EIC-CDR, ref-EIC-white}.

\section{Materials and Methods}

\subsection{Physics of Stage I}

In the initial stage NICA will provide proton beams up to 5 GeV with collision luminosity up to $10^{31} cm ^{-2} s^{-1}$ for the $pp$ collisions and up to 4.5 GeV/n (per nucleon) deuteron beams with collision luminosity up to $10^{30} cm ^{-2} s^{-1}$ for the first few years. There are also possibilities of asymmetric collisions like $pd$ as well as light nuclei (i.e. $C, Ca$) collisions.

The SPD will take advantage of the low energies at the initial stage to look for compelling and interesting physics effects in $pp$, $dd$ and possibly in the light nuclei collisions. Various physics goals and programs for this initial stage are discussed in detail in the published work \cite{ref-stage1}.

\subsubsection{Spin effects in elastic collisions} 
Measurements of the $pp$ elastic scattering cross-sections in small angles ($\theta \sim 3-10^{\circ}$) will access a kinematic region of momentum transfer $|t| \sim 0.1 - 0.8$ GeV$^2$. Small oscillations in the $t$-dependence probe the proton structure involving mesons in the periphery (pion cloud model). The SPD will provide high precision data in this region to test the models of two-pion exchange process in the elastic scattering. 

Glauber models with Gribov inelastic corrections have been successful in describing elastic $dd$ scattering data at a few tens of GeV. At the first stage energies of up to $\sqrt{s} = 9$ GeV/n, unpolarized $dd$ cross-section measurements and subsequent comparisons with calculations will test if the inelastic corrections are relevant for this kinematic regime.

At large angles $\theta \sim 90^{\circ}$, $dd \rightarrow dd$ processes are sensitive to the six-quark structure of the deuterons. The SPD will make cross-section measurements from $dd$ elastic collisions at large $\theta_{CM}$ to search for non-nucleonic degrees of freedom. 

\subsubsection{Charmonium production}
The SPD will measure light and charm meson productions near the production threshold. Of particular interest is the charmonium ($J/\Psi$) formation near threshold for $pp$ and $dd$ collisions as it will test the isotopic dependence (involvement of protons or neutrons) on the production due to different spin structure of the corresponding matrix elements.

Threshold production of charmonia in ion-ion collisions is also considered as a promising probe of the quark-gluon plasma (QGP).

\subsubsection{Strange hypernuclei production}
Although there has been no evidence of stable hypernuclei of baryon number $A = 2$, there are measurements \cite{ref-hypernuc} of candidates ($^3_{\Lambda}He, ^3_{\Lambda}H$) with baryon number $A = 3$. There have been proposals to look for neutral hypernucleus $^4_{\Lambda \Lambda}n$ in the $dd$ collisions at the SPD. Calculations predict a peak in the production at $\sqrt{s} = 5.2$ GeV. A measurement of this hypernuclei with strangeness $S = -2$ would be the first of its kind.

\subsubsection{Other interesting physics at stage I}
Measurements during the stage I of the SPD will also test various effects that can be broadly categorized as multi-quark correlations. These include nuclear PDFs involving fluctons or multi-quark degrees of freedom, higher twist contributions of two or three quark correlations in PDFs, multi-parton scattering in hadronic and nuclear collisions and formation of exotic multi-quark resonance states (i.e. tetraquark and pentaquark).

\subsection{Physics of Stage II}
For stage II when NICA will reach its full potential of luminosity ($10^{32} cm ^{-2} s^{-1}$ for the $pp$ collisions), energy and polarization capacities, the SPD will focus primarily on making measurements of observables from polarized $pp$ and $dd$ collisions that are sensitive to the gluon distributions inside nucleons. Detailed discussions of the access to gluon contents from the measurements at the SPD can be found in the article \cite{ref-stage2}. At peak luminosities, one year of data at the SPD will correspond to integrated luminosities of $1.0$ and $0.1 \ \rm fb^{-1}$ respectively for p-p collisions at $\sqrt{s} = 27$ and $13.5$ GeV.

\begin{figure}[H]
\begin{subfigure}[h]{0.5\textwidth}
 \includegraphics[width=\textwidth]{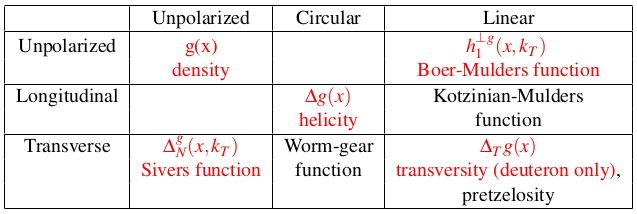}
 \caption{\centering  }
 \label{fig1a}
\end{subfigure}
\hfill
\begin{subfigure}[h]{0.5\textwidth}
 \includegraphics[width=\textwidth]{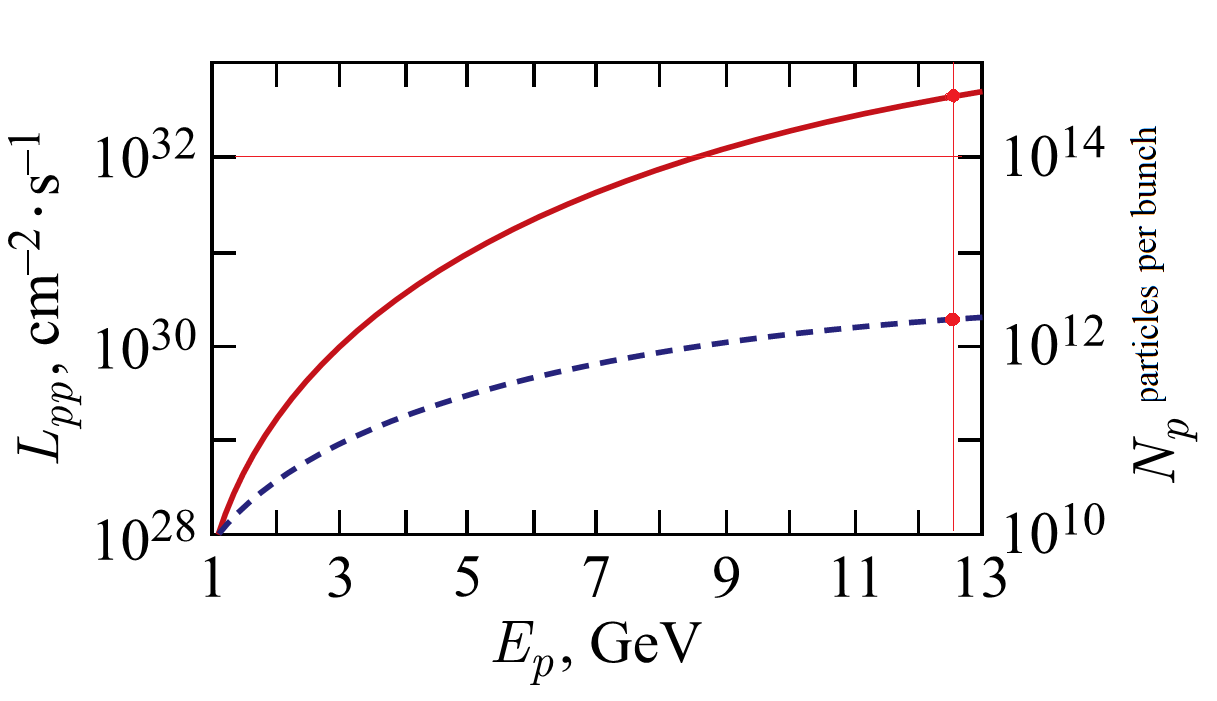}
 \caption{\centering  }
 \label{fig1b}
\end{subfigure}
\caption{(a) PDFs in red (color online) will be accessed in measurements at the SPD. (b) Expected luminosity, energy and bunch intensity for proton beams at NICA.}
\label{fig1}
\end{figure}

Measurements of asymmetries and correlations from the polarized proton-proton collisions at the SPD will, in particular, be sensitive to gluon helicity, Sivers and Boer-Mulders distributions. Measurements from the polarized deuteron collisions will access gluon transversity and tensor polarized gluon distribution inside deuterons. NICA will be the first facility to provide polarized deuteron beams in such energy range and the SPD will have the unique ability to access quantities that have not been measured before.

Unpolarized cross-section measurements at the SPD will provide data sensitive to the unpolarized gluon distributions ($g(x)$). Double-helicity asymmetry measurements ($A_{LL}$) at the SPD will probe gluon helicity distribution function($\Delta g(x)$), single transverse spin asymmetries ($A_N$) will provide access to the gluon Sivers function ($\Delta^g_N(x, k_T)$) and measurements of the azimuthal correlations of hadron pair production from unpolarized $pp$ collisions will probe the Boer-Mulders distributions ($h^{\perp}_1(x,k_T)$). Double and single vector/tensor asymmetries from polarized $dd$ collisions will respectively probe the gluon transversity ($\Delta g_T(x)$) and tensor polarized gluon PDF ($C^T_G(x)$).

\subsection{Detectors for Stage I}

The SPD detector system \cite{ref-TDR} will have complete $4\pi$ coverage in solid angle. The design includes a barrel part and two end-caps. In the barrel part, the SPD will feature a solenoid magnet providing a field up to 1.2 T at the interaction point. The magnetic field provides charge separation of the particle tracks and also helps in determination of the charged particle momentum.

\begin{figure}[h]
\centering
\includegraphics[width=0.85\textwidth]{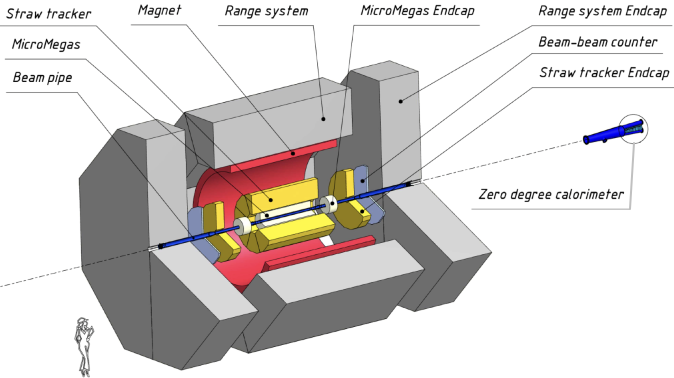}
\caption{\centering Schematic of the SPD detector at stage I.\label{fig2}}
\end{figure}  

Going outward from the aluminium beam-pipe, the detectors in the barrel part of the SPD at this stage will include :
\begin{enumerate}
 \item Micromegas tracker that will help charged particle momentum reconstruction.
 \item Multi-layer tracker system with PET (metal coated polyethylene terephthalate) straws arranged along Z,U,V (U,V are stereo layers at $5^{\circ}$ with Z straws along the beam direction) with a spatial resolution $\sim 150 \ \mu m$. Tracker will provide charged particle momentum as well as limited particle identification using energy depositions $-\frac{dE}{dx}$ in the straw layers with an energy resolution $\frac{\delta E}{E} = 8.5\%$.
 \item A range system (RS) just outside the magnet consisting of layers of mini Drift Tubes (MDT) and absorbing material ($Fe$). RS will provide muon-to-hadron separation of the charged tracks and hadronic calorimetry.
\end{enumerate}
End-caps of the the SPD detector system at stage I will consist of : micromegas, straw tracker, beam-beam counter (BBC) that will provide local polarimetry, luminosity control and collision timing information, range system and zero-degree calorimeter (ZDC) in far forward and backward positions that will provide local polarimetry, luminosity control and event selection criteria for elastic collisions.

\subsection{Detectors for Stage II}

For the second stage of the operations, due to different requirements of the physics in focus at this stage, some parts will be replaced and new detectors will be included \cite{ref-TDR}.

\begin{figure}[h]
\centering
\includegraphics[width=\textwidth, height=0.5\textwidth]{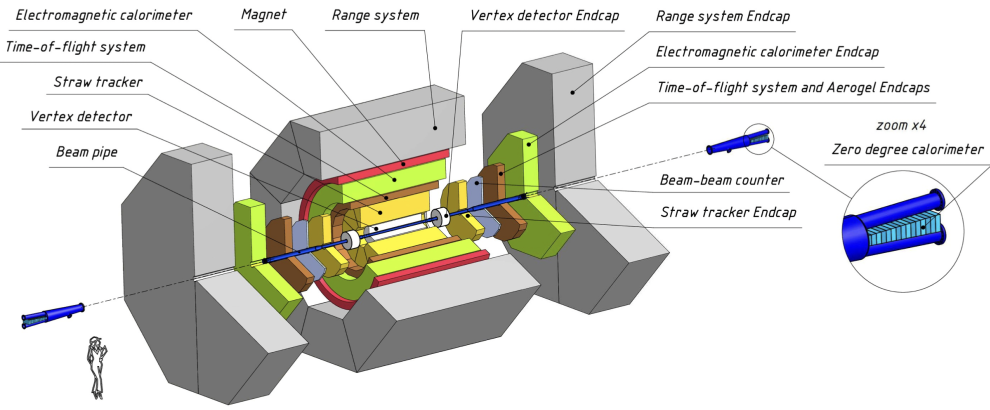}
\caption{\centering Schematic of the SPD detector at stage II.\label{fig3}}
\end{figure}

For stage II, the barrel part of the SPD will consist of :
\begin{enumerate}
 \item An improved silicon vertex detector to replace the Micromegas from stage I. Two options being considered are (1) monolithic active pixel sensor (MAPS) and (2) double silicon strip detector (DSSD). The new component will contribute to tracking, momentum determination and specifically in reconstructing secondary vertices for the decays of short lived particles. MAPS silicon tracker will provide a secondary vertex position resolution of $40-60 \ \mu m$.
 \item Straw tracker. Tracking system will provide a momentum resolution $\frac{\delta p_T}{p_T} = 2\%$ for 1 GeV/c momentum tracks (same resolution in stage I).
 \item Time-of-flight (TOF) detector for particle identification with a timing resolution of $50$ ps and $\pi/K$ separation for charged tracks up to 1.5 GeV/c momentum.
 \item Electromagnetic calorimeter for the determination of photon energies with an energy resolution $\frac{\delta E}{E} = \frac{5\%}{\sqrt{E}} \oplus 1\%$ and electron/positron identification .
 \item Range system.
\end{enumerate}
The endcaps will also have some new components : silicon vertex detector, straw tracker, BBC, TOF detector, Aerogel detector for extending the $\pi/K$ separation up to 2.5 GeV/c momentum, electromagnetic calorimeter and ZDC.

\subsection{Detector Performance}

\begin{figure}[h]
\begin{subfigure}[h]{0.31\textwidth}
 \includegraphics[width=\textwidth]{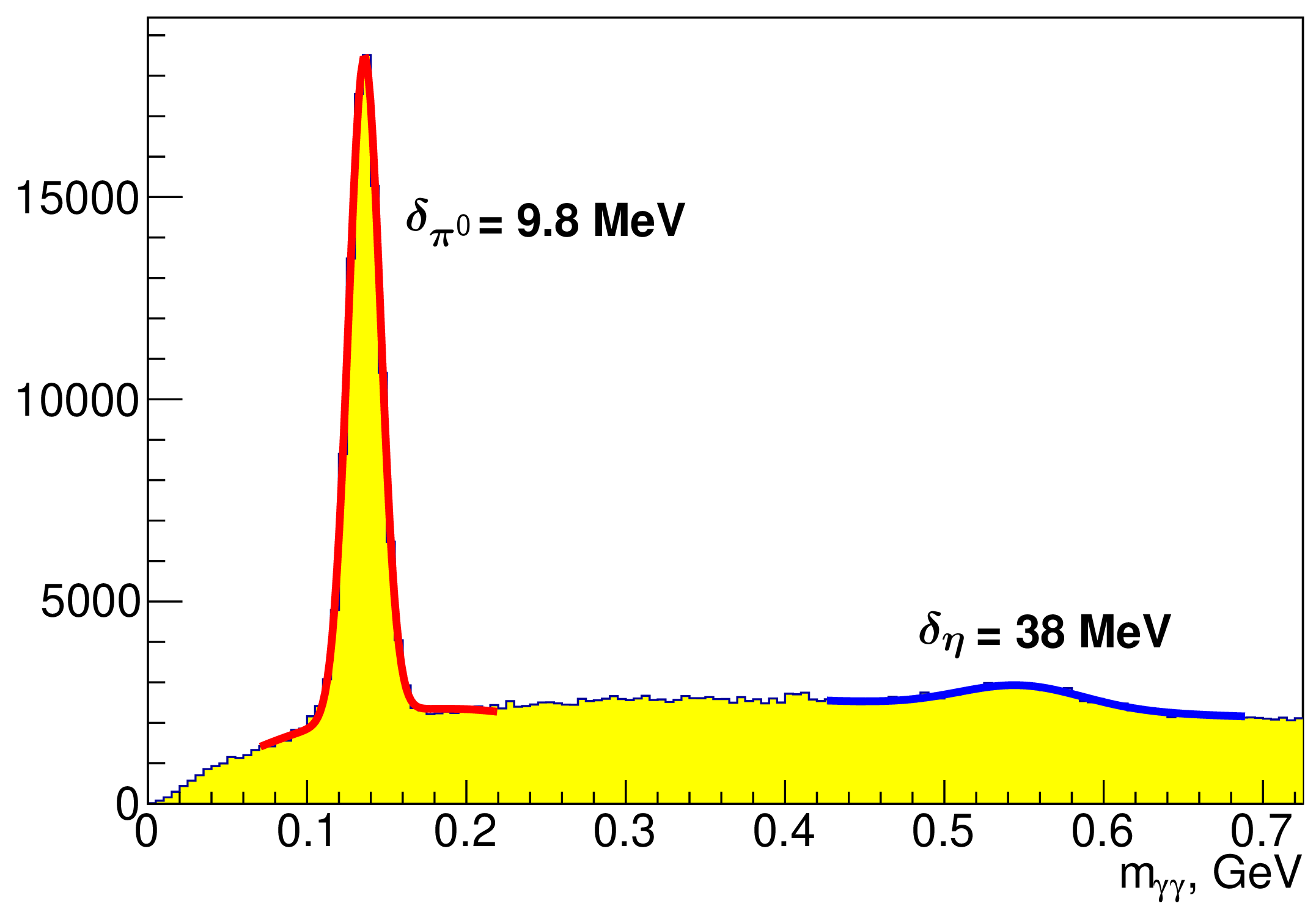}
 \caption{\centering  }
 \label{fig9a}
\end{subfigure}
\hfill
\begin{subfigure}[h]{0.36\textwidth}
 \includegraphics[width=\textwidth]{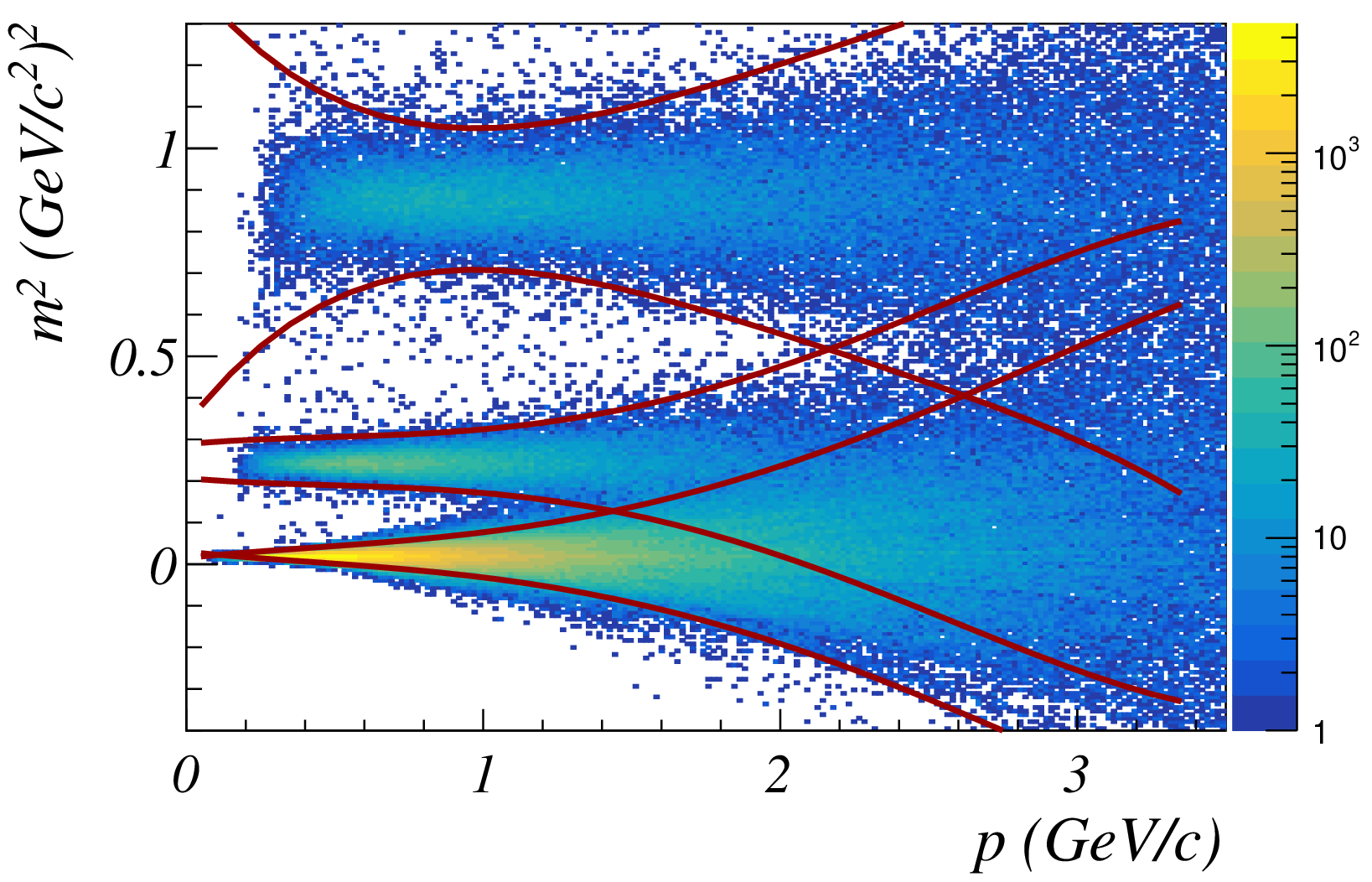}
 \caption{\centering  }
 \label{fig9b}
\end{subfigure}
\hfill
\begin{subfigure}[h]{0.31\textwidth}
 \includegraphics[width=\textwidth]{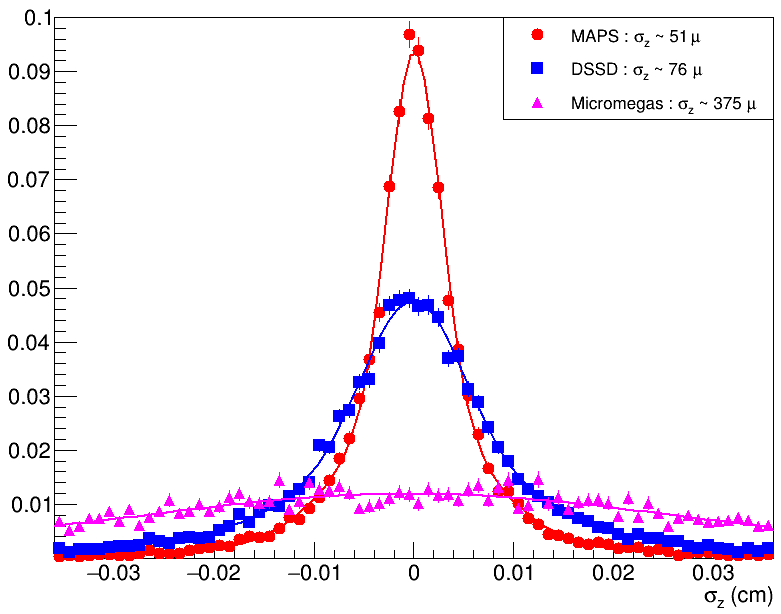}
 \caption{\centering  }
 \label{fig9c}
\end{subfigure}
\caption{(a) Mass resolution of pion mass reconstruction from two photons. (b) Mass-squared vs. momentum at the time-of-flight detector. (c) $D^0$ secondary vertex resolution along beam direction.}
\label{fig9}
\end{figure}

Figure (\ref{fig9}) shows Monte Carlo simulation performance of some of the detectors to be used in key measurements. From the left, in Figure (\ref{fig9a}) two photon invariant mass spectra using the electromagnetic calorimeter shows the pion mass resolution ($\delta_m = 9.8$ MeV) that can be achieved. Figure (\ref{fig9b}) illustrates the particle identification using the time-of-flight detector. Pion-kaon separation can be achieved for particle momentum up to 1.5 GeV/c. Figure (\ref{fig9c}) illustrates the secondary vertex resolution along the beam direction for three possibilities of central tracking detectors, namely micromegas for the first stage and DSSD or MAPS for the second stage. MAPS based detectors is clearly the best performing detector providing a secondary vertex position resolution of $\delta_z \sim 50 \ \mu$m.

\section{Results}

In order to access various gluon distributions, the SPD will focus on processes involving gluonic interactions. Three major channels of interest at the the SPD are :
\begin{itemize}
 \item {\bf Quark-gluon scattering to prompt photons}. This is a particularly clean channel for theoretical interpretations as it does not involve hadronization.
 \item {\bf Gluon fusion to charmonia production}($J/\Psi, \Psi(2S), \chi_{c1/c2}$). Measurements at the SPD will be primarily via di-muon decay channels of the charmonia.
 \item {\bf Gluon fusion to open-charm mesons}. $D$ mesons at the SPD will be detected via hadronic decay channels. This is the highest statistics channel but a challenging measurement due to large amount of combinatorial background.
\end{itemize}

\begin{figure}[H]
\begin{subfigure}[h]{0.325\textwidth}
 \includegraphics[width=\textwidth]{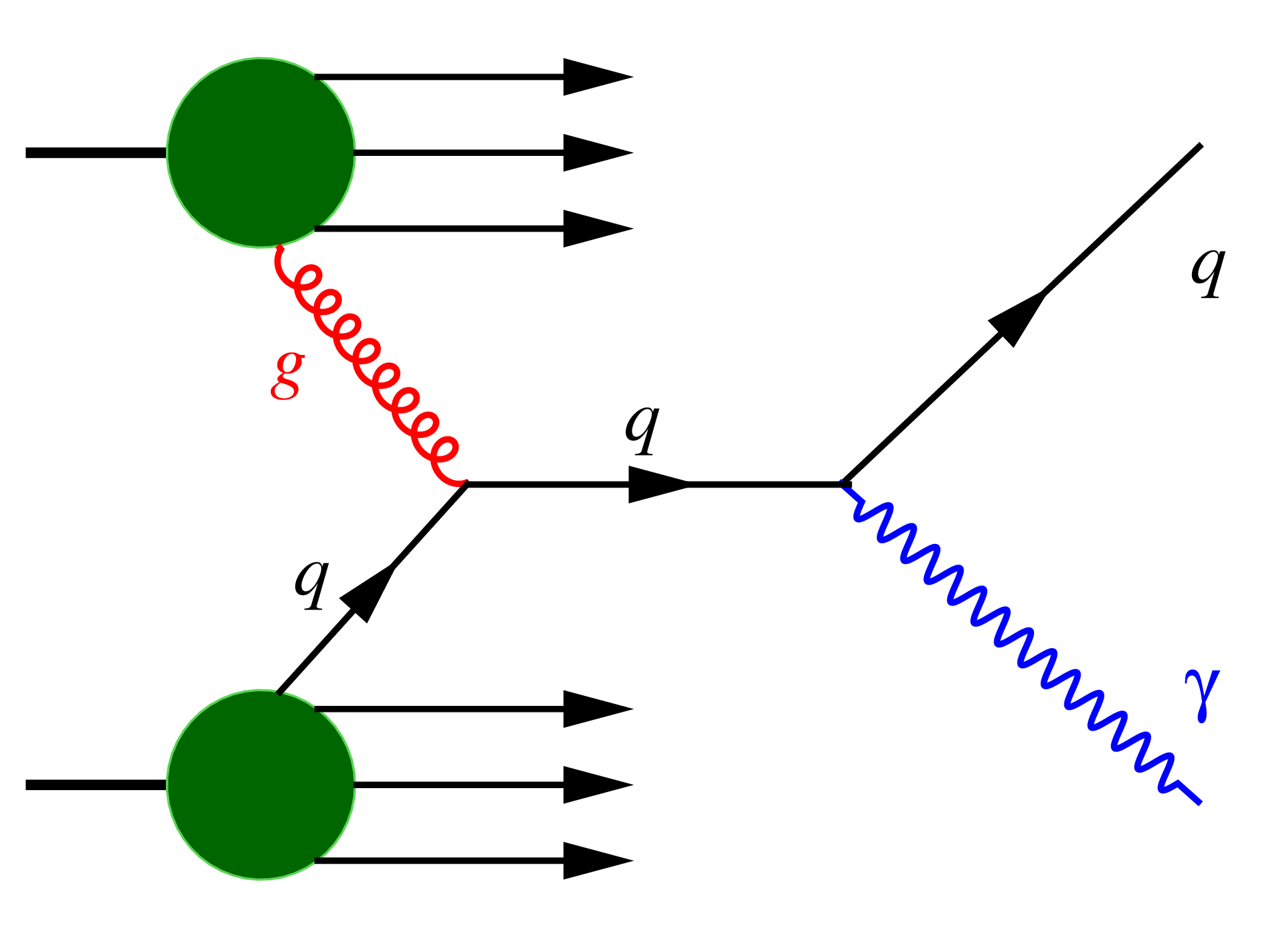}
 \caption{\centering  }
 \label{fig11b}
\end{subfigure}
\hfill
\begin{subfigure}[h]{0.325\textwidth}
 \includegraphics[width=\textwidth]{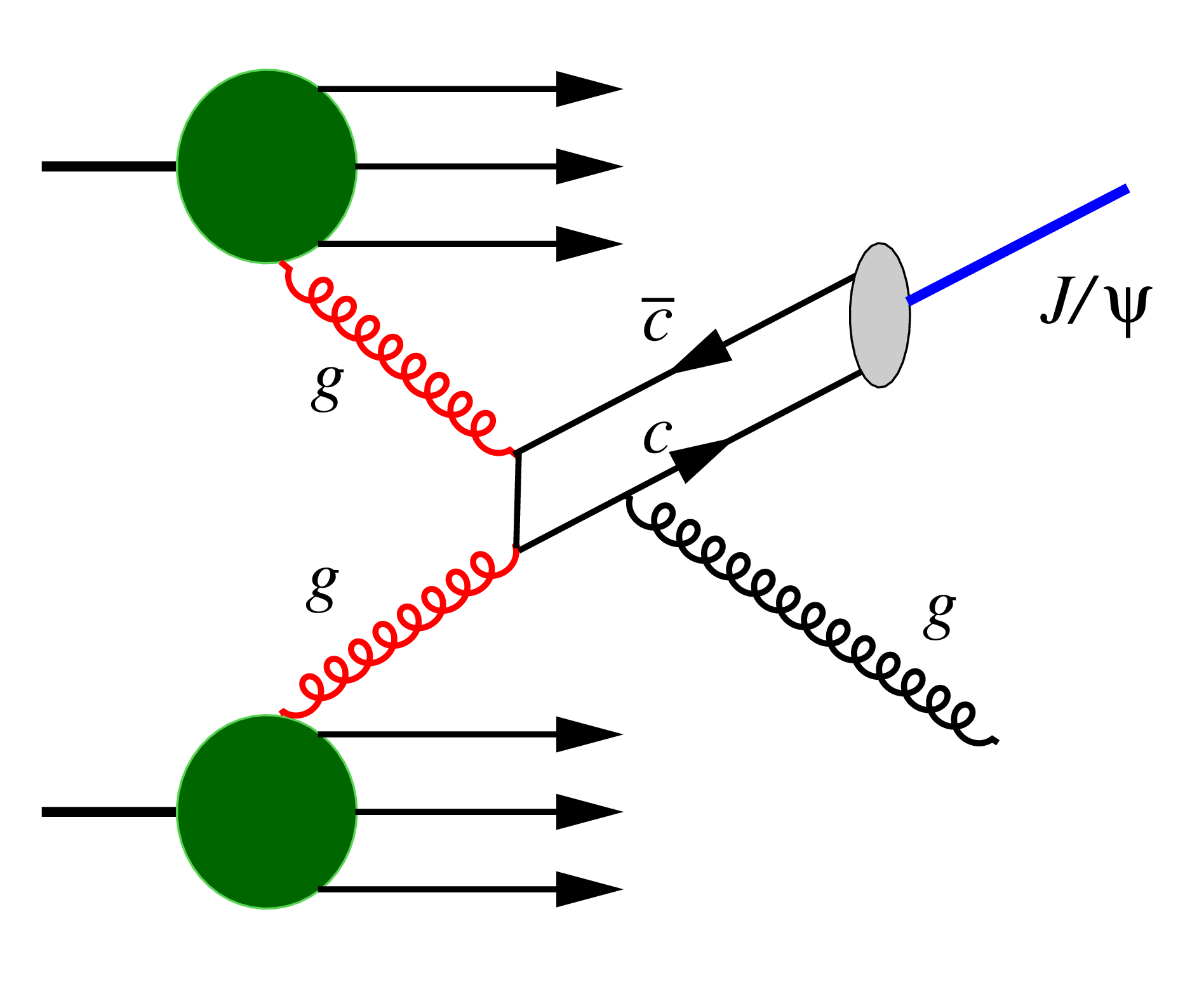}
 \caption{\centering  }
 \label{fig11a}
\end{subfigure}
\hfill
\begin{subfigure}[h]{0.325\textwidth}
 \includegraphics[width=\textwidth]{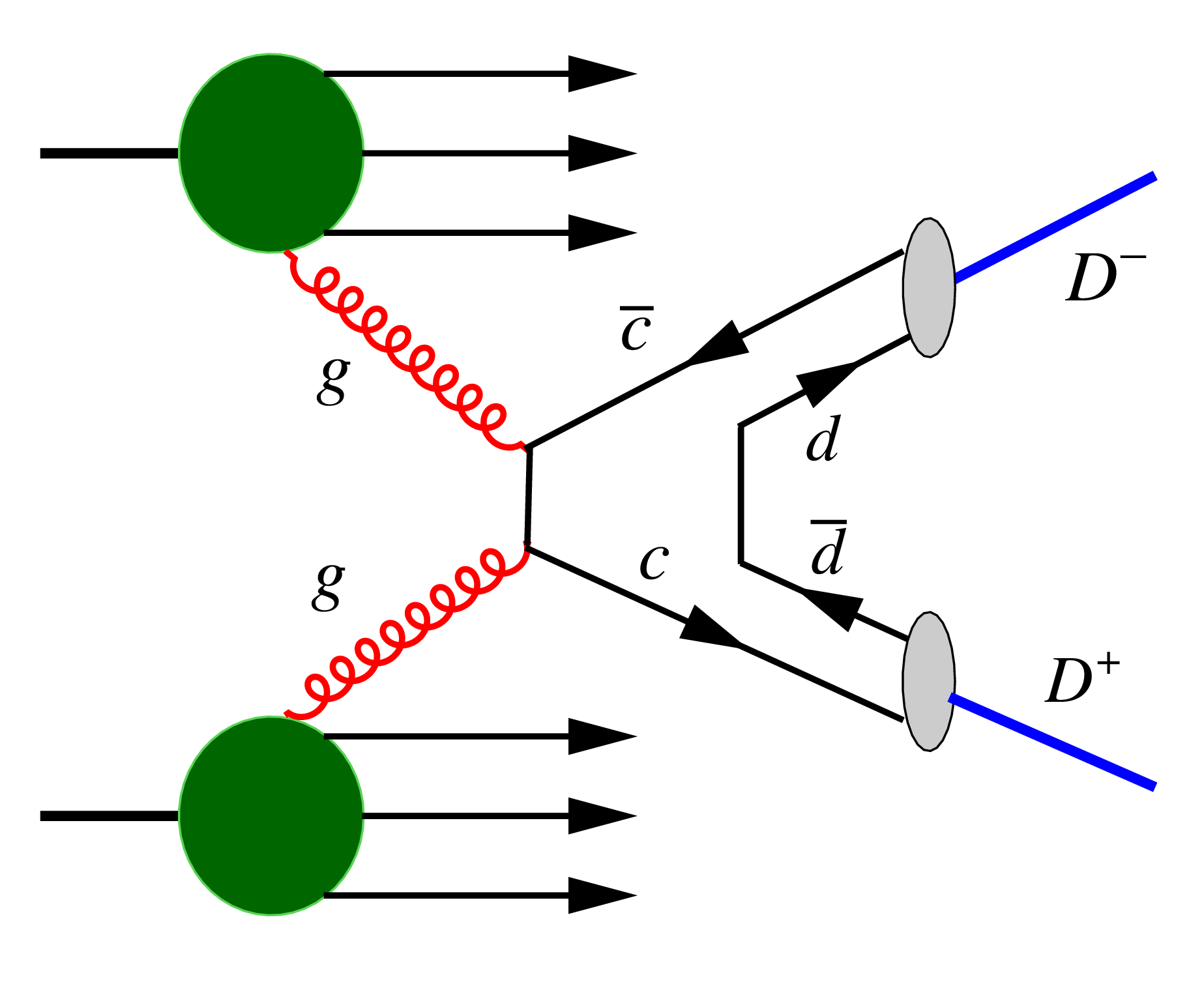}
 \caption{\centering  }
 \label{fig11c}
\end{subfigure}
\caption{Schematics of partonic sub-processes of interest : (a) quark-gluon scattering to prompt photon production (b) gluon fusion to charmonia production (c) gluon fusion to open charm production.}
\label{fig11}
\end{figure}

\begin{figure}[h]
\centering
\includegraphics[width=0.45\textwidth]{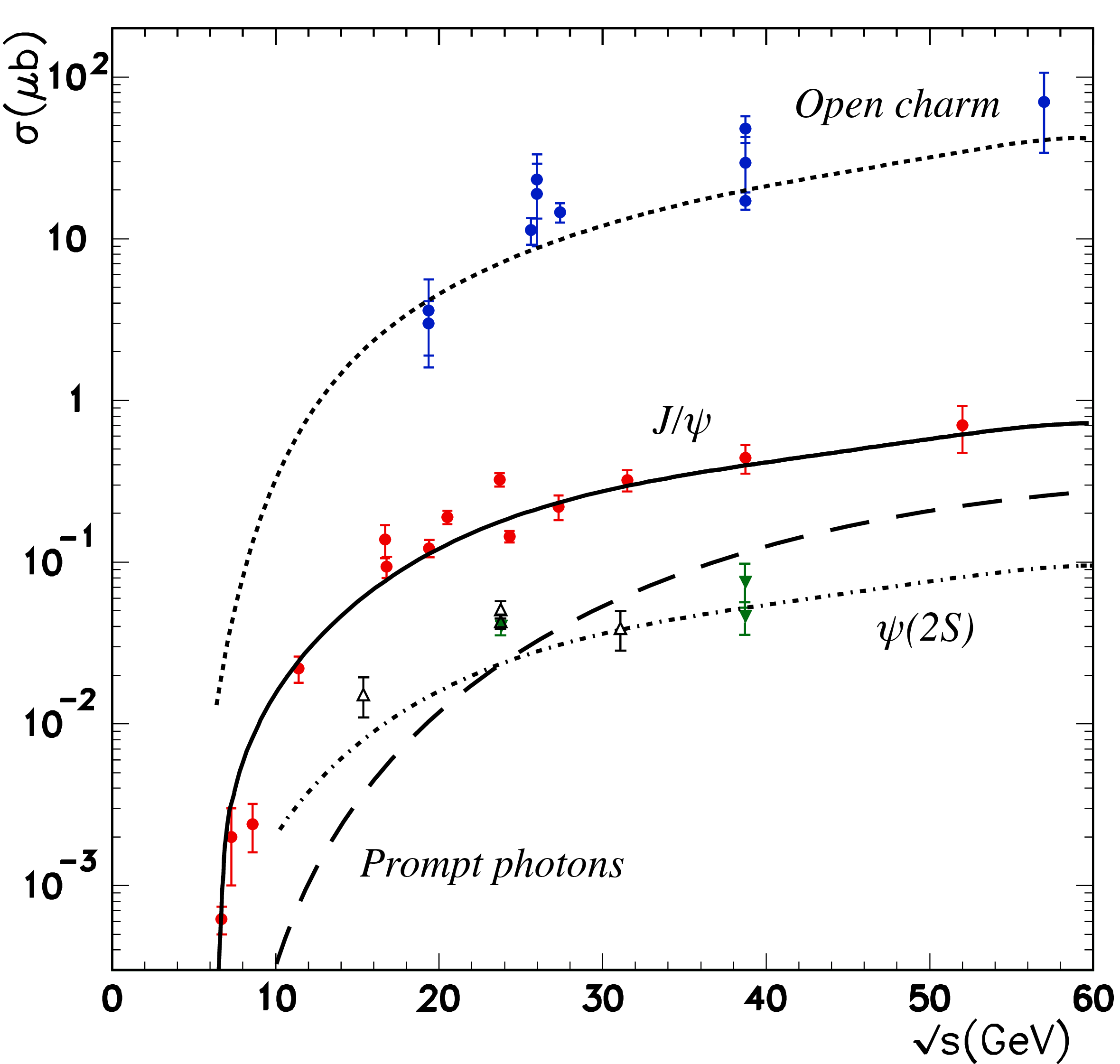}
\caption{\centering Cross-sections of three channels of interest at SPD kinematics \cite{ref-cHadroprod}. \label{fig10}}
\end{figure}

As illustrated above (Figure \ref{fig10}), at the peak SPD energy of $\sqrt{s} = 27$ GeV, open charm processes are the most abundant among these three. However, the hadronic channels of charmed meson decays are notoriously difficult to measure because of the orders of magnitudes more combinatorial background arising from random combinations of hadrons from other hard processes. Charmonia are comparatively easier to measure via di-muon decay channels with good muon-hadron separation (using Range System at the SPD). Prompt photons, while the rarest among these processes, has the advantage of being the cleanest probe for theoretical interpretations. This channel also requires careful estimation of background arising from decays of light neutral mesons ($\pi^0, \eta$).

\subsection{Prompt Photon Measurements}

Prompt photon productions in the leading order may occur via gluon Compton scattering and quark-antiquark annihilation. However, at SPD energies, the $q\bar{q} \rightarrow g\gamma$ contribution is small. Fragmentation contribution (from scattered (anti-)quarks) to the prompt photon production is also estimated to be small (15-30 \%) \cite{ref-stage2} making prompt photons an excellent tool to probe gluon distributions inside nucleons. Measurements will be made using the electromagnetic calorimeter and the photons from neutral light meson (i.e. $\pi^0, \eta$) decays are the largest source of background. Untagged photons from $\pi^0$ can be estimated using MC simulations.

\begin{figure}[h]
 \centering
 \includegraphics[width=0.5\textwidth]{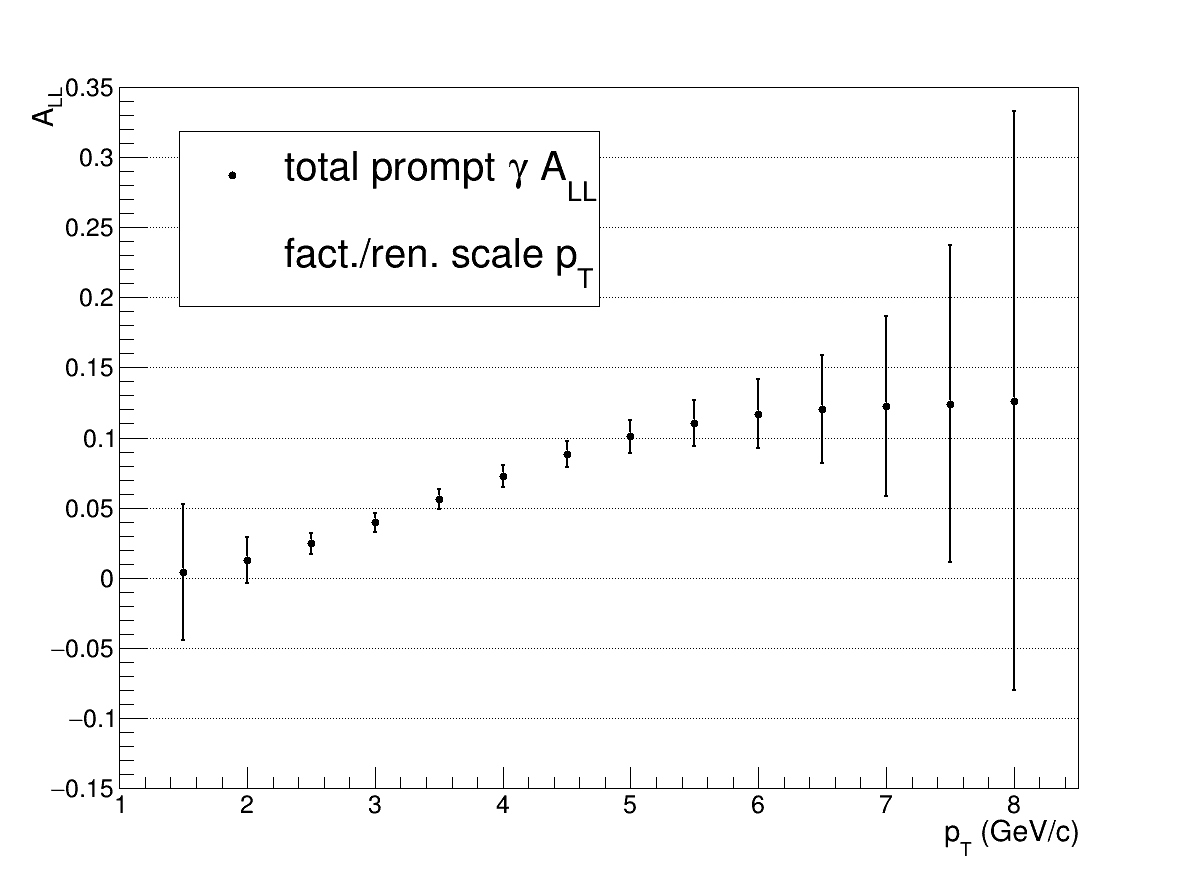}
 \caption{Prompt photon double helicity asymmetry as function of transverse momentum calculated using NNPDF3.0 unpolarized and DSSV2014 polarized PDF with projected uncertainties from measurements with one year of data at the SPD.}
 \label{fig7a}
\end{figure}

\begin{figure}[H]
 \centering
 \includegraphics[width=0.75\textwidth]{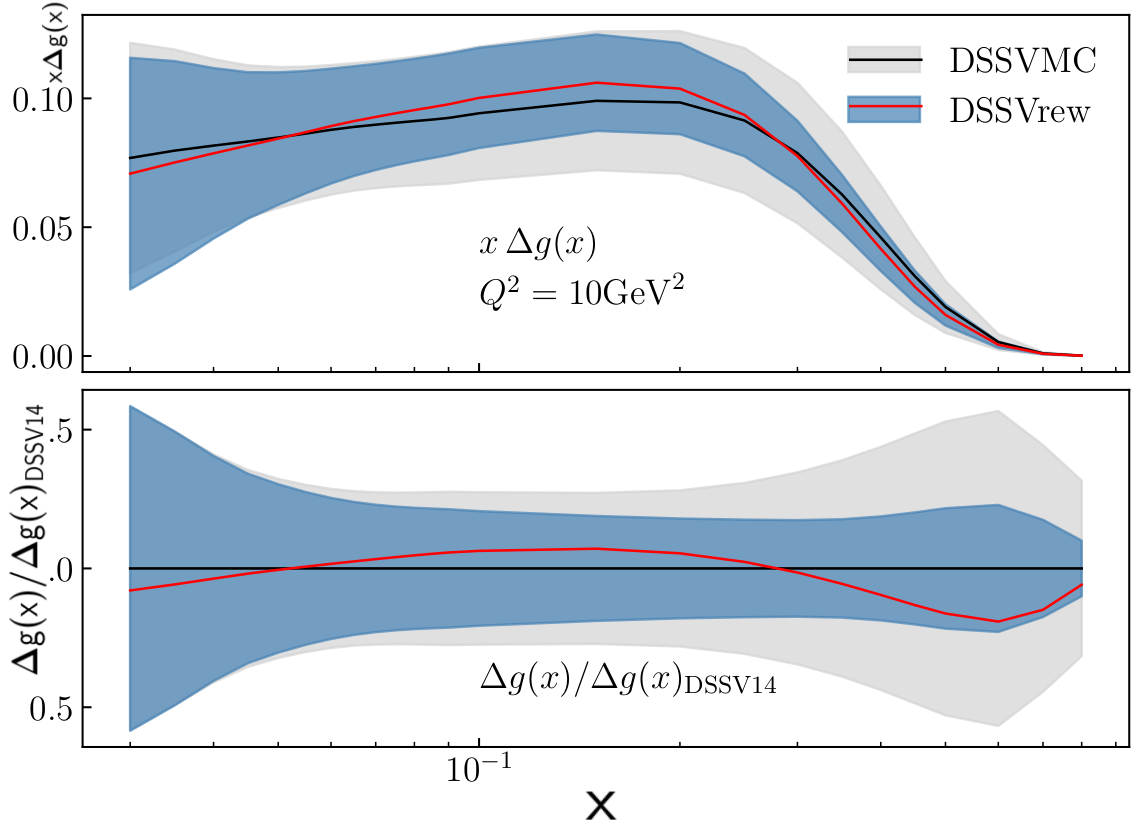}
 \caption{Estimated impact of $A^{\gamma}_{LL}$ measured at the SPD. Black and red solid lines are respectively the mean of the thousand replicas of the DSSV2014 polarized gluon PDF before and after the re-weighting using the projected SPD measurements. Light blue and grey bands respectively are the spread of replicas indicating the statistical uncertainties.}
 \label{fig7b}
\end{figure}

Double helicity asymmetry of the prompt photons at the SPD is sensitive to the gluon distribution in the high momentum fraction range. Recent works \cite{ref-DSSVrewt} have tested inclusion of new measurements with Monte Carlo re-weighting instead of full extraction of PDFs, creating an efficient technique to estimate the impact of new data points on a global analysis of PDF extraction. A similar study \cite{ref-DSSVpvt}  using the NNPDF3.0 unpolarized and DSSV2014 as the polarized PDF set estimates the impact of the $A^{\gamma}_{LL}$ measurement with the projected statistical uncertainties (Figure \ref{fig7a}) for one year of recorded data at the SPD. Figure \ref{fig7b} illustrates that in the high-x region ($0.2 \leq x \leq 0.8$) SPD measurements can be used to reduce the uncertainties of the gluon helicity distribution ($\Delta g(x)$) by a factor of $\sim 2$.

\begin{figure}[h]
\begin{subfigure}[h]{0.5\textwidth}
\centering
 \includegraphics[width=0.9\textwidth]{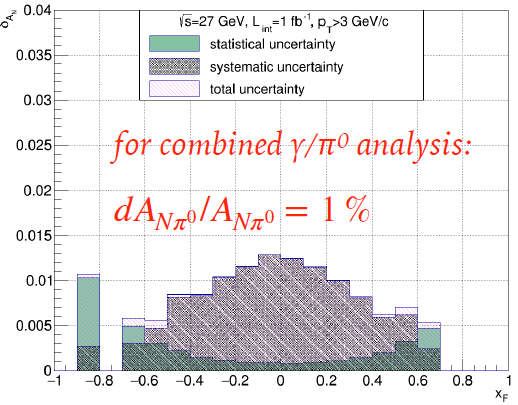}
 \caption{\centering  }
 \label{fig8a}
\end{subfigure}
\hfill
\begin{subfigure}[h]{0.5\textwidth}
 \includegraphics[width=\textwidth]{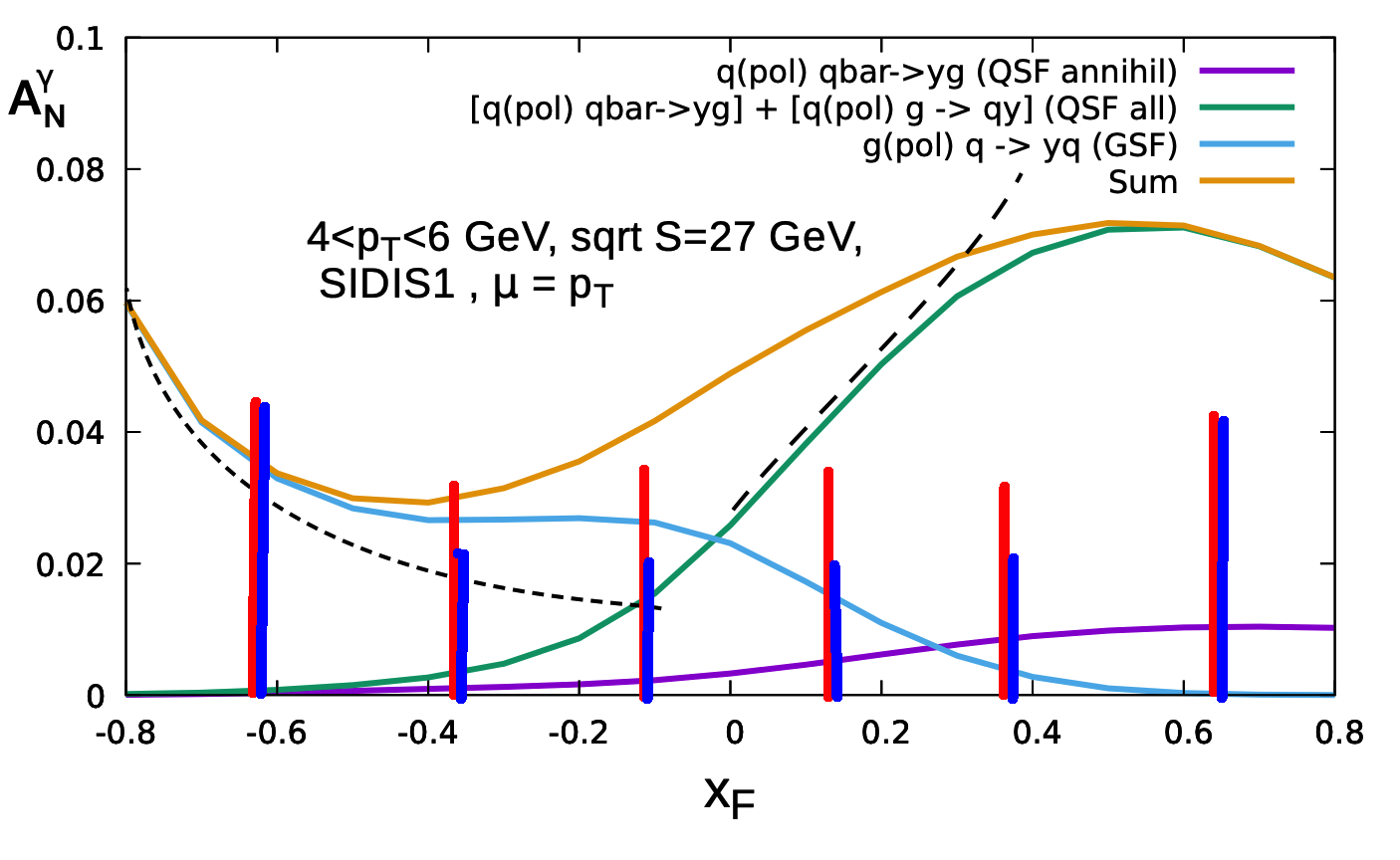}
 \caption{\centering  }
 \label{fig8b}
\end{subfigure}
\caption{(a) Estimated uncertainties for $A^{\gamma}_N$ as function of $x_F$. (b) Theoretical estimates of $A^{\gamma}_{N}$ at the SPD. Calculations performed using SIDIS1 \cite{ref-GSF-SIDIS1} parameterization of the Gluon Sivers Function.}
\label{fig8}
\end{figure}

Theoretical estimates of the single transverse spin asymmetries show (Figure \ref{fig8b}) that the asymmetries in the forward $x_F$ region are dominated by the quark-antiquark annihilation process whereas the gluon dominated process generates asymmetries in the backward $x_F$ region.

\subsection{Charmonia Measurements}

Charmonia production at the SPD energies ($10-27$ GeV) is dominated by the gluon-gluon fusion process \cite{ref-stage2}. Charmonia measurement via di-muon invariant mass spectra using the Range System as muon identifier is a powerful tool at the SPD. Mass resolution of $\sim 40$ MeV or better is expected for $J/\Psi$ from di-muon invariant mass spectra.

\begin{figure}[H]
\begin{subfigure}[h]{0.328\textwidth}
 \includegraphics[width=\textwidth]{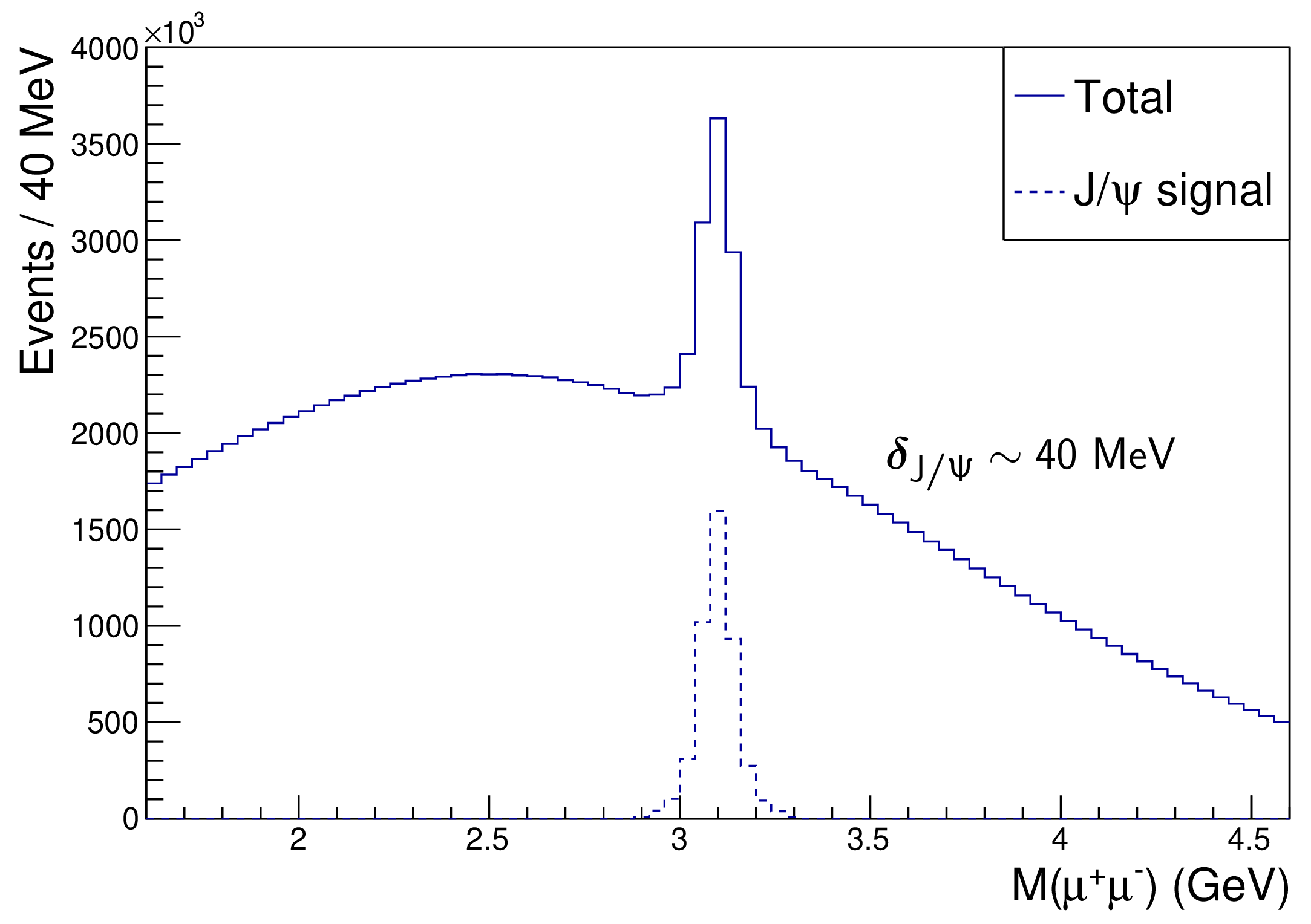}
 \caption{\centering  }
 \label{fig4a}
\end{subfigure}
\hfill
\begin{subfigure}[h]{0.328\textwidth}
 \includegraphics[width=\textwidth]{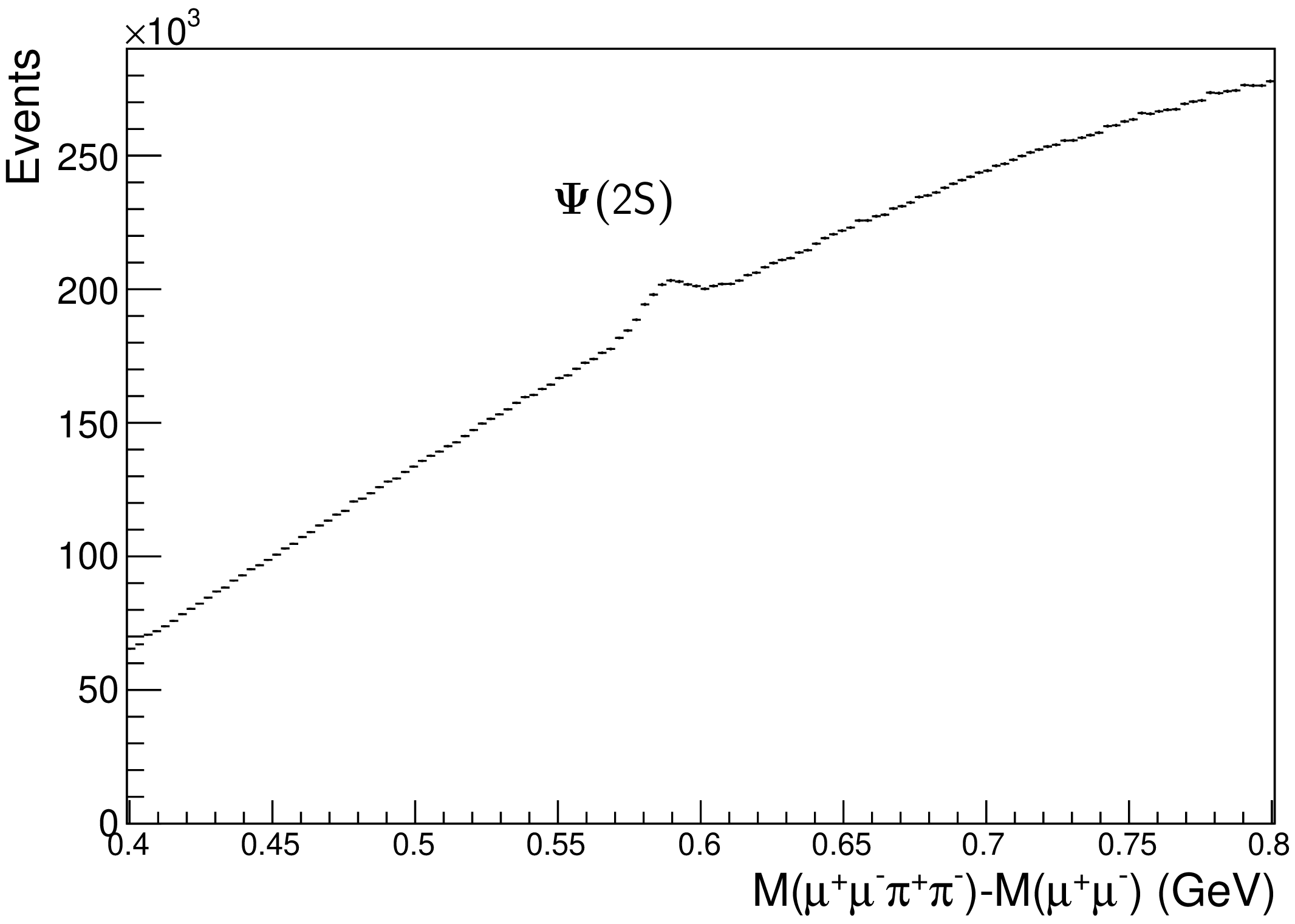}
 \caption{\centering  }
 \label{fig4b}
\end{subfigure}
\hfill
\begin{subfigure}[h]{0.328\textwidth}
 \includegraphics[width=\textwidth]{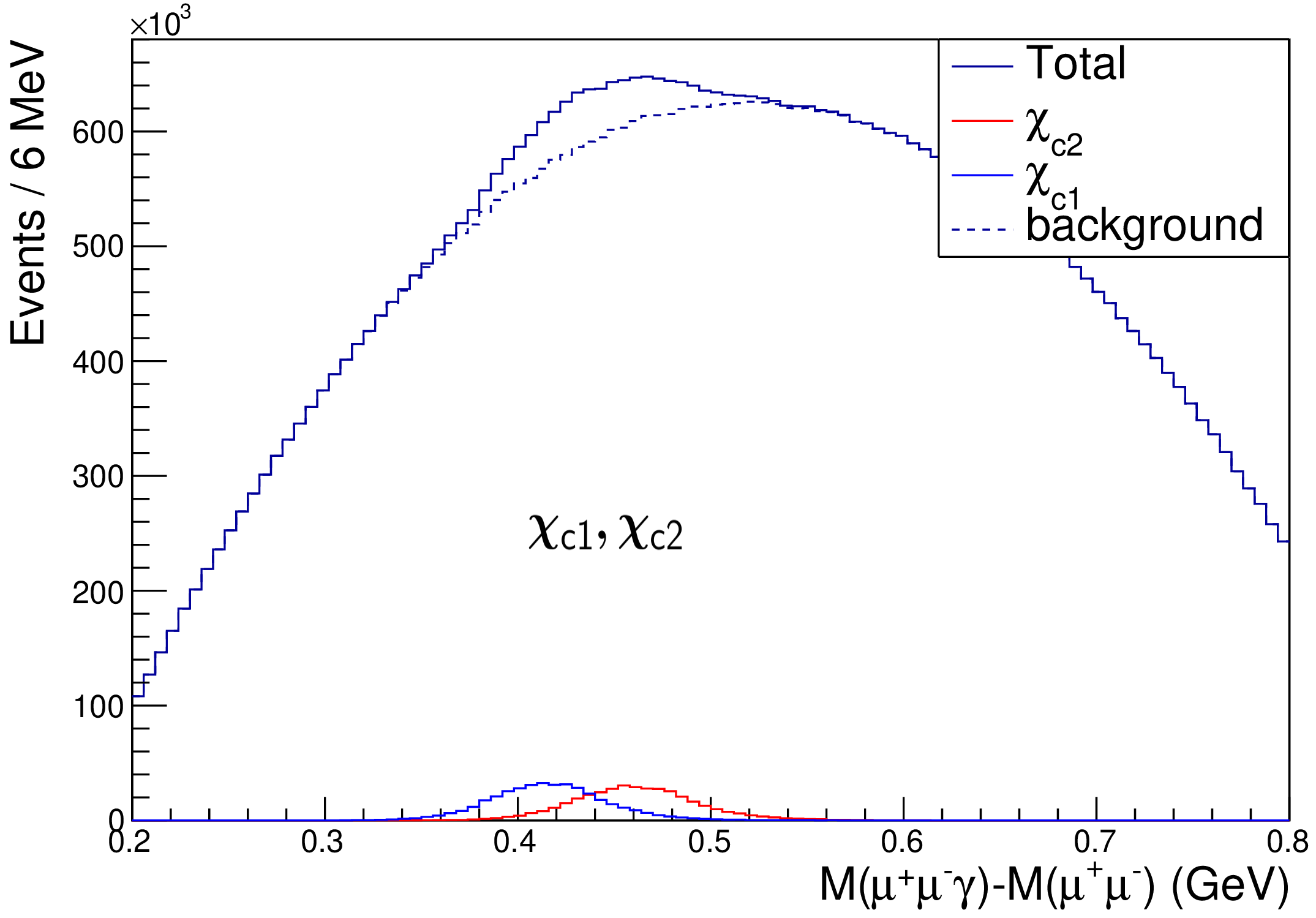}
 \caption{\centering  }
 \label{fig4c}
\end{subfigure}
\caption{From Monte Carlo simulation studies at the SPD : (a) Di-muon invariant mass spectra for $J/\Psi$ measurements. (b) Invariant mass spectra for $\Psi(2S)$ measurements. (c) Invariant mass spectra showing $\chi_{c1/c2}$ peaks.}
\label{fig4}
\end{figure}

 About $\sim 12$ M events with $J/\Psi$ are expected from one year of data at peak luminosity at the SPD \cite{ref-stage2}. It will also be possible to perform measurements of rarer charmonia. $\Psi(2S)$ can be detected via $\mu^+\mu^-$ and $\mu^+\mu^-\pi^+\pi^-$ decay channels. About seven hundred thousand events producing $\Psi(2S)$ are expected in one year of data at the SPD. Moreover, $\chi_{c1}$ and $\chi_{c2}$ can also be measured via $\gamma \mu^+\mu^-$ channel and about 2.5 M events including both types are expected in one year of data at the peak luminosity.

\begin{figure}[h]
\begin{subfigure}[h]{0.5\textwidth}
 \includegraphics[width=0.95\textwidth]{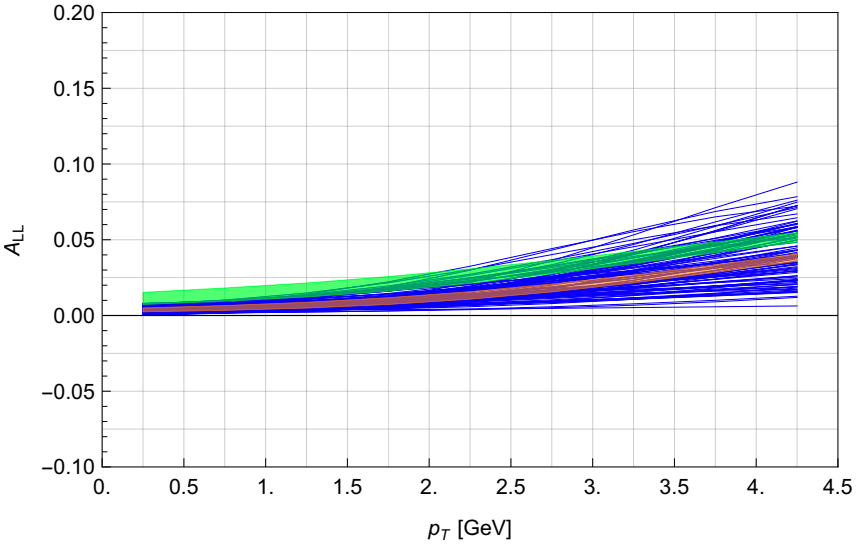}
 \caption{\centering  }
 \label{fig5a}
\end{subfigure}
\hfill
\begin{subfigure}[h]{0.5\textwidth}
 \includegraphics[width=\textwidth]{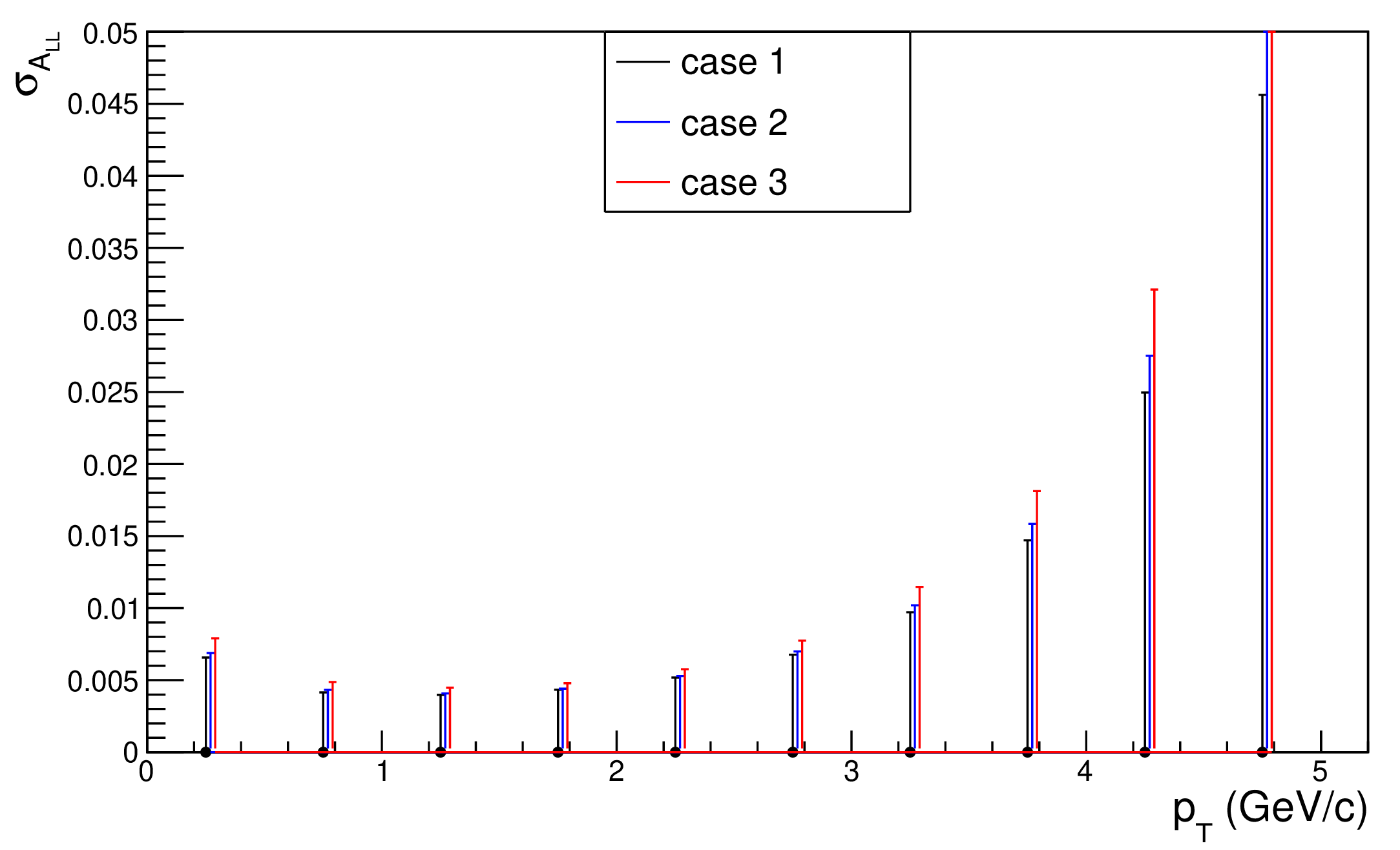}
 \caption{\centering  }
 \label{fig5b}
\end{subfigure}
\caption{(a) Estimated $A^{J/\Psi}_{LL}$ as function of $p_T$ calculated using unpolarized NNPDF NLO and polarized NNPDFpol1.1 sets. The green band indicates uncertainties due to hadronization parameters (LDME) and the brown band indicates scale uncertainties. (b) Projected statistical uncertainties of $A^{J/\Psi}_{LL}$ for three different cases of selection cuts on the muon polar angles are shown.}
\label{fig5}
\end{figure}

Alongside unpolarized cross-section, which can be used to compare with theoretical estimations to shed light on the poorly understood hadronization models of charmonia, double helicity asymmetry ($A_{LL}$) and single transverse spin asymmetry ($A_N$) will also be measured. It will be possible to perform high precision $A^{J/\Psi}_{LL}$ measurements (Figure \ref{fig5b}) at the SPD that will improve on the current knowledge of the gluon helicity PDF. 

\begin{figure}[H]
\centering
\includegraphics[width=0.8\textwidth]{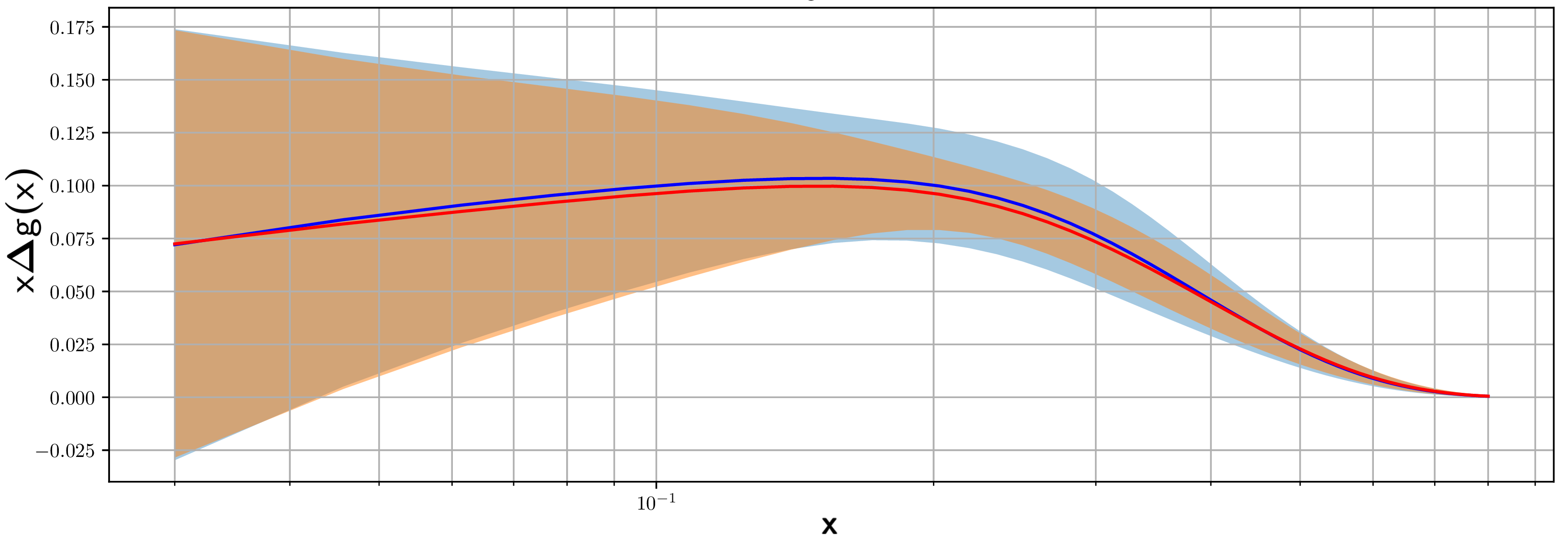} \\
\includegraphics[width=0.8\textwidth]{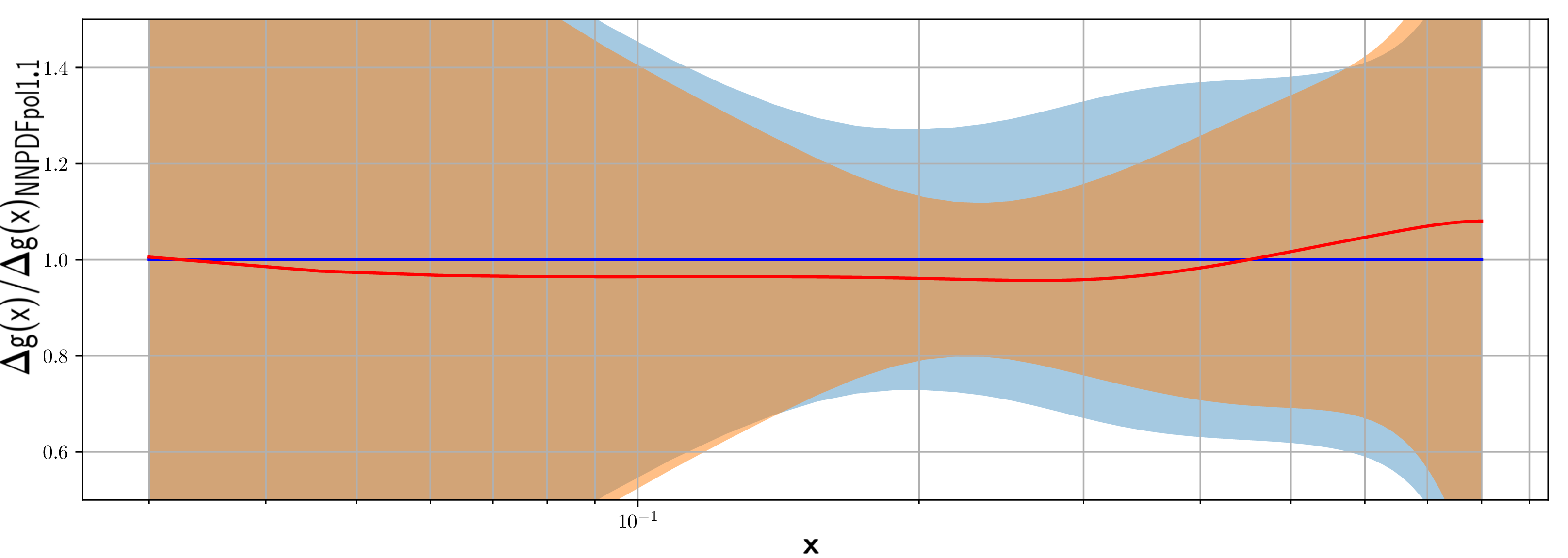}
\caption{Impact of $A^{J/\Psi}_{LL}$ measurements at the SPD. Blue and red lines respectively show the mean of the NNPDFpol1.1 replica sets before and after the re-weighting using the projected uncertainties of SPD measurements. Light blue and light orange bands respectively show the spread of hundred replicas of the PDF before and after the re-weighting. \label{fign6}}
\end{figure}

In a recent simulation study, $A^{J/\Psi}_{LL}$ were calculated using NNPDF NLO unpolarized and NNPDFpol1.1 polararized sets. Using a technique similar to \cite{ref-DSSVrewt}, the resulting calculations of $A^{J/\Psi}_{LL}$ were used with the projected uncertainties at the SPD for one year of data to re-weight the polarized PDF replicas to estimate the impact of the measurement at the SPD. Figure \ref{fign6} illustrates the impact of $J/\Psi$ double helicity asymmetry measurements at the SPD. Expected measurements will reduce uncertainties in the Bjorken-x range of $0.2 \leq x \leq 0.5$ significantly.

\begin{figure}[H]
\begin{subfigure}[h]{0.5\textwidth}
 \includegraphics[width=\textwidth,height=0.75\textwidth]{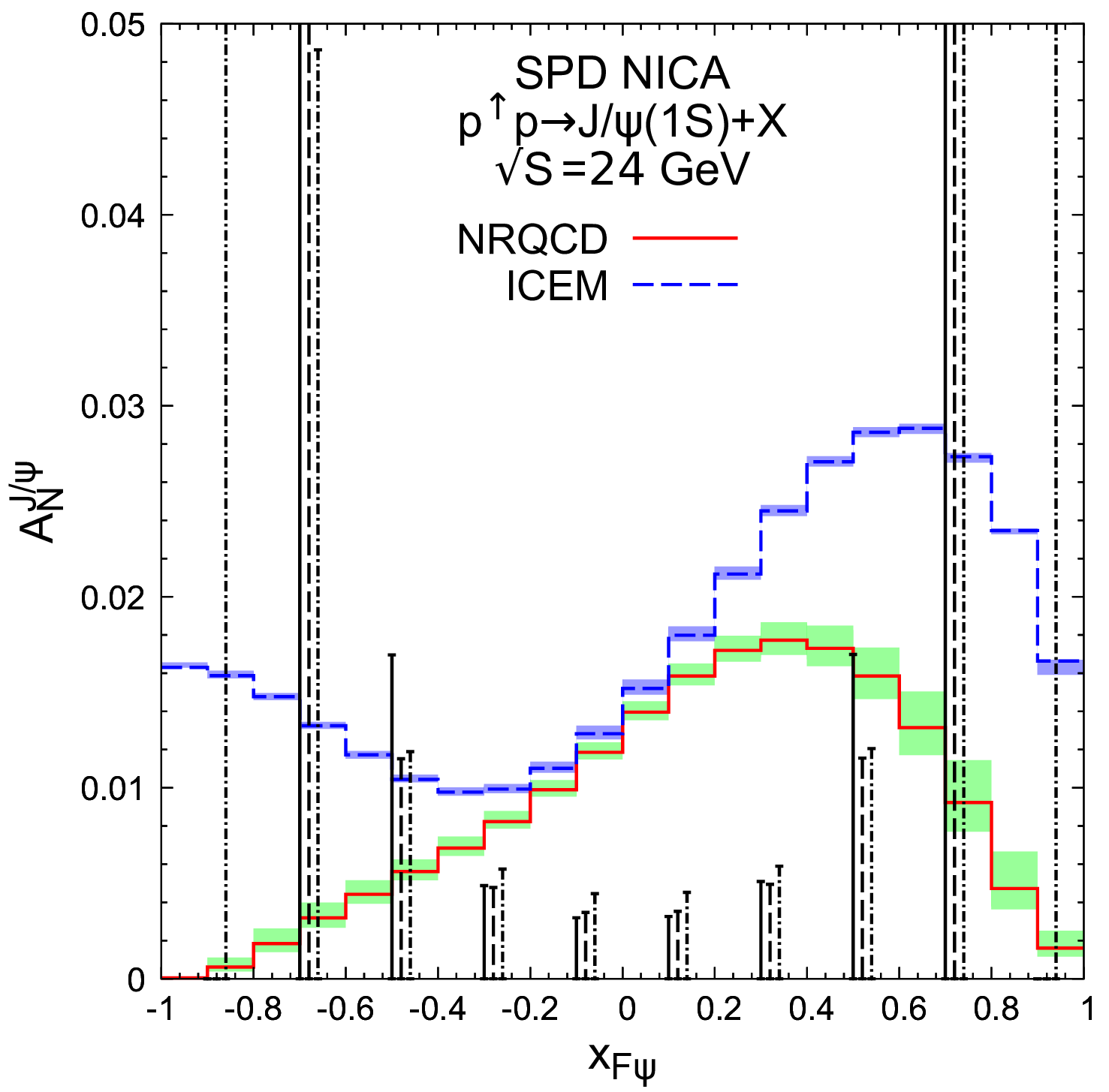}
 \caption{\centering  }
 \label{fig5a}
\end{subfigure}
\hfill
\begin{subfigure}[h]{0.5\textwidth}
 \includegraphics[width=\textwidth,height=0.75\textwidth]{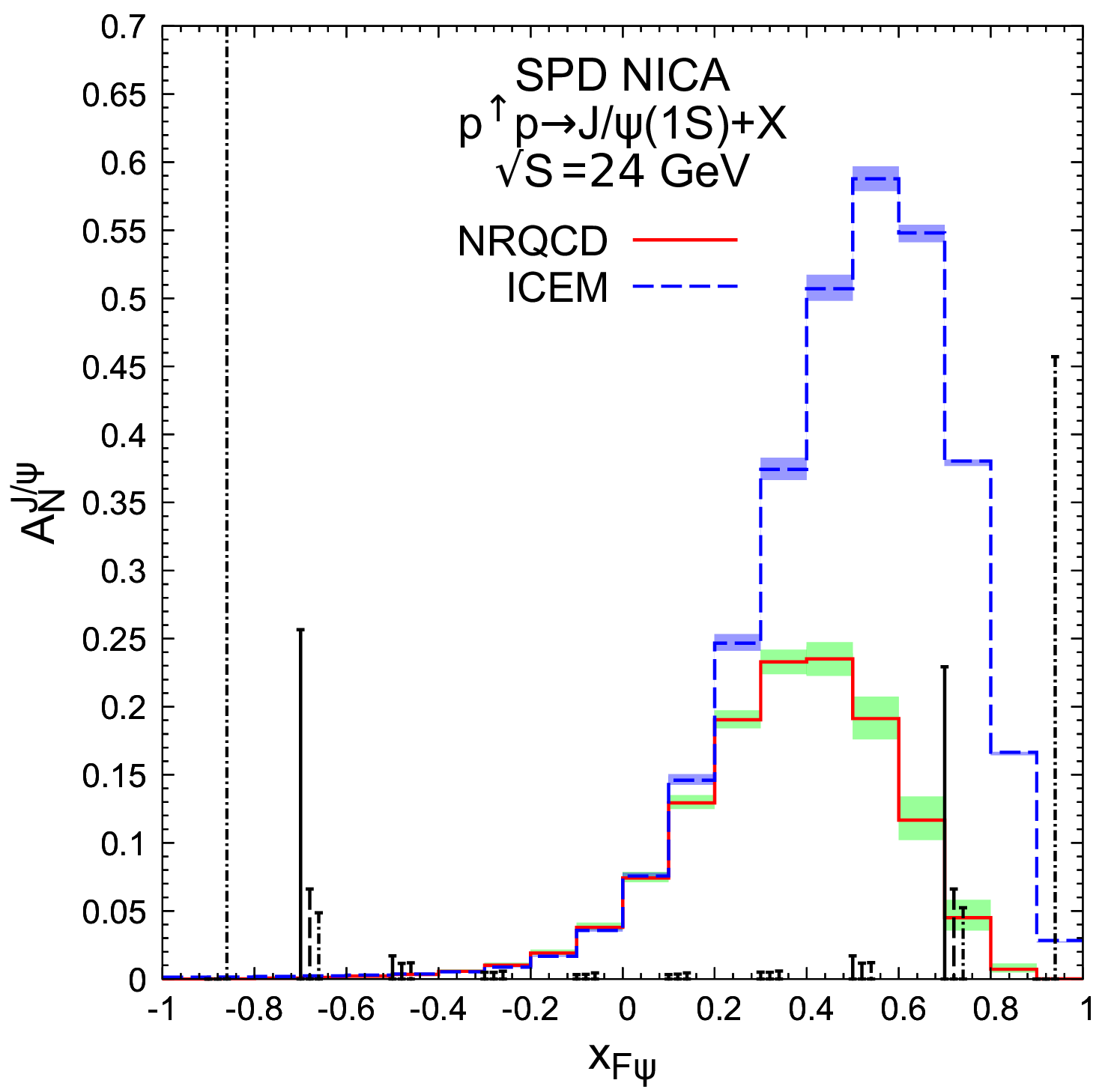}
 \caption{\centering  }
 \label{fig5b}
\end{subfigure}
\caption{Estimated single transverse spin asymmetries of $J/\Psi$ as function of $x_F$ using generalized parton model (GPM) with two different parameterizations ((a) D'Alesio \cite{ref-GSF-Dalesio} and (b) SIDIS1 \cite{ref-GSF-SIDIS1}) of the Gluon Sivers Function. Projected statistical uncertainties (for three different selection cuts on muon polar angles) for one year of data at the SPD are displayed.  \label{fig6}}
\end{figure}

Transverse single spin asymmetries of $J/\Psi$ are sensitive to the gluon Sivers distributions. Theoretical estimations \cite{ref-jpsi-AN} of the $A^{J/\Psi}_N$ depend very strongly on the choice of the parton models and hadronization models and estimations can differ by an order of magnitude depending on the phenomenological parameterization used as shown in Figure \ref{fig6}. High precision measurements of $A^{J/\Psi}_N$ can be extremely useful in reducing such model dependence in this kinematic regime.

\subsection{Open Charm Measurements}

Open charm meson spin asymmetries have been measured in DIS experiment like COMPASS \cite{ref-COMPASS-openC} to estimate gluon polarization but it has not been measured as yet in $pp$ collider experiments. At the SPD, open charm D mesons will be detected through their hadronic decay channels i.e. $D^0 \rightarrow \pi^+ K^-$, $D^+ \rightarrow \pi^+ \pi^+ K^-$ and their antiparticle counterparts. Figure (\ref{fig10}) shows that the open charm production cross-sections are almost two orders of magnitude larger than the charmonium production cross-sections, making them quite abundant at the SPD kinematics. However, the abundance of charged pions and kaons from other hard scattering processes make it a particularly challenging measurement. The combinatorial background from pions and kaons from other processes is more than four orders of magnitude larger than the signal.

\begin{figure}[H]
\begin{subfigure}[h]{0.5\textwidth}
\includegraphics[width=\textwidth, height=0.85\textwidth]{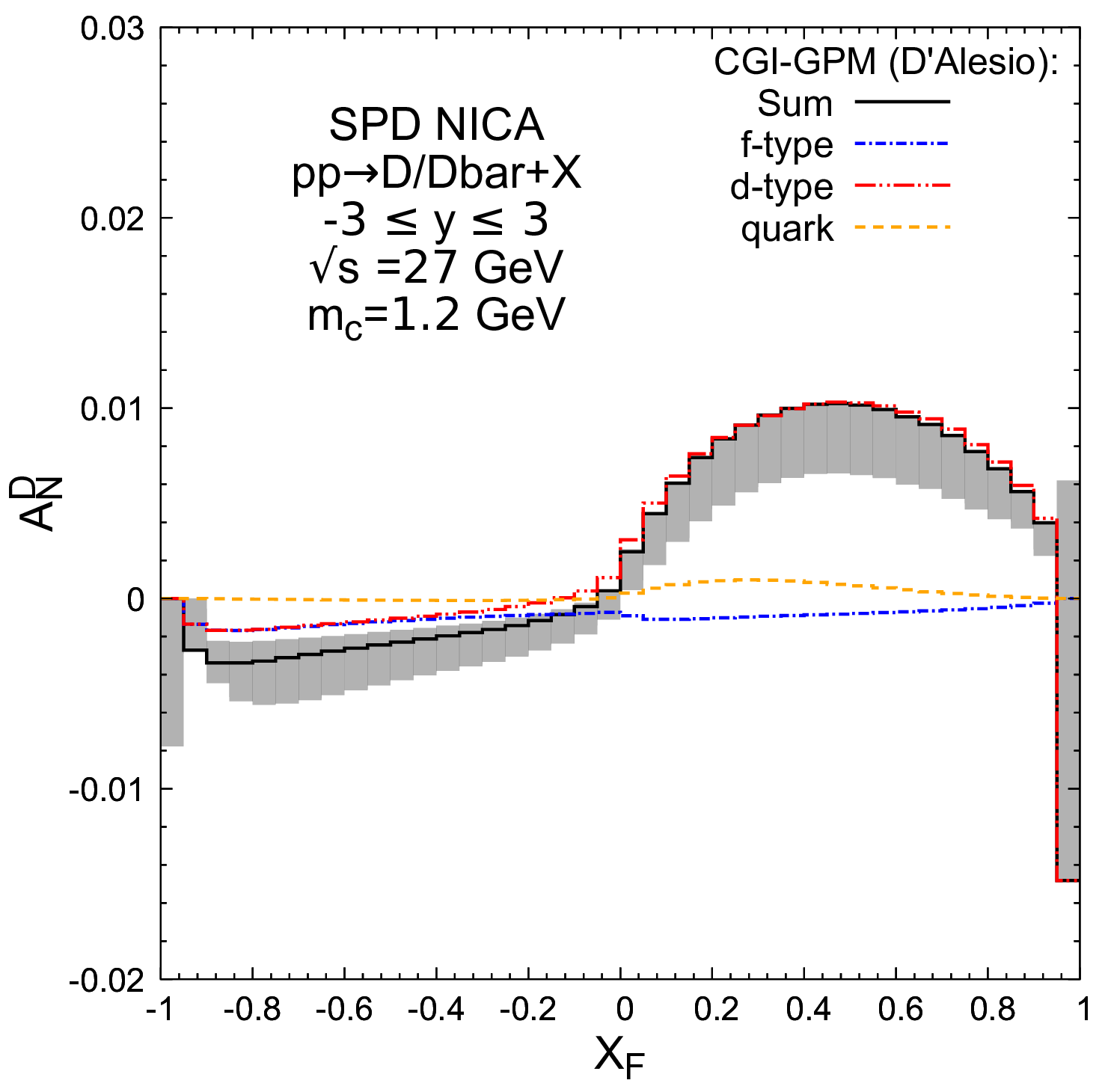}
\caption{\centering  }
 \label{fig13a}
\end{subfigure}
\hfill
\begin{subfigure}[h]{0.5\textwidth}
\includegraphics[width=\textwidth, height=0.85\textwidth]{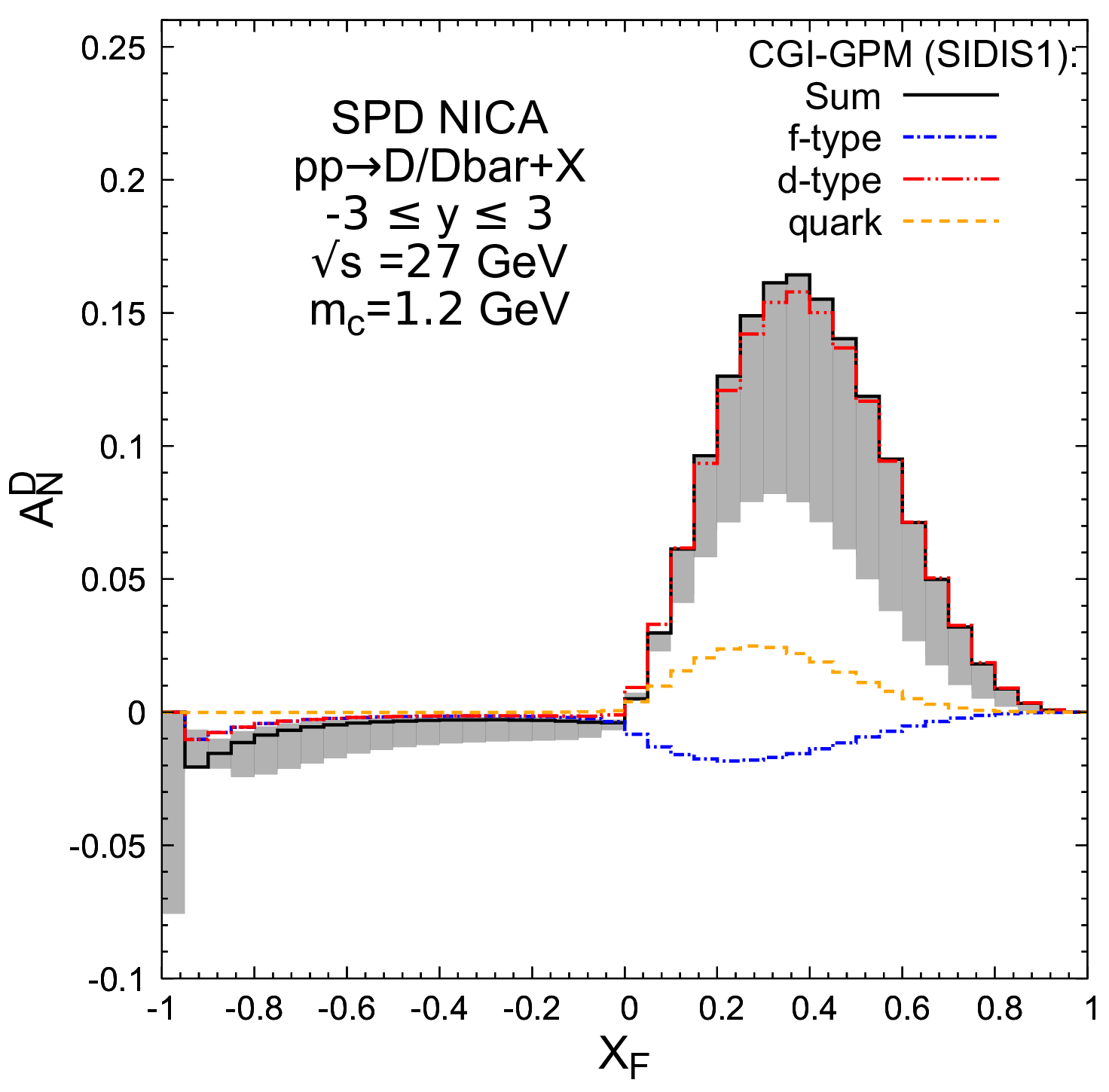}
\caption{\centering  }
 \label{fig13b}
\end{subfigure}
\caption{(a) Estimated transverse single spin asymmetry of inclusive D meson productions at the SPD using color-gauge-invariant generalized parton model (CGI-GPM). The first two panels (a) and (b) show the dependence on different sets of parameters of the Gluon Sivers Function. (c) 
\label{fig13}}
\end{figure}

Theoretical calculations of the transverse single spin asymmetry for inclusive D mesons at the SPD kinematics (Figure \ref{fig13}) using the color gauge invariant generalized parton model (CGI-GPM) show significant expected asymmetries in the forward region ($x_F > 0.2$) whereas for backward $x_F$ the asymmetry is compatible with zero. However, the size of the asymmetry depends strongly on the parameterization used for the parton model.

\begin{figure}[H]
\begin{subfigure}[h]{0.5\textwidth}
 \includegraphics[width=\textwidth]{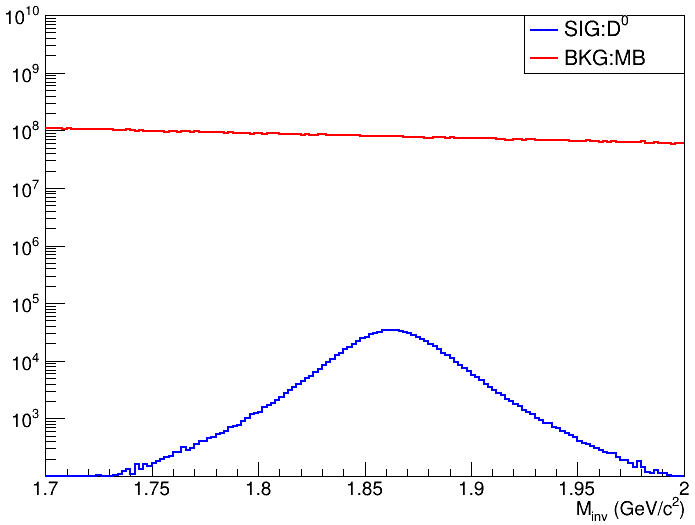}
 \caption{\centering  }
 \label{fig12a}
\end{subfigure}
\hfill
\begin{subfigure}[h]{0.5\textwidth}
 \includegraphics[width=\textwidth]{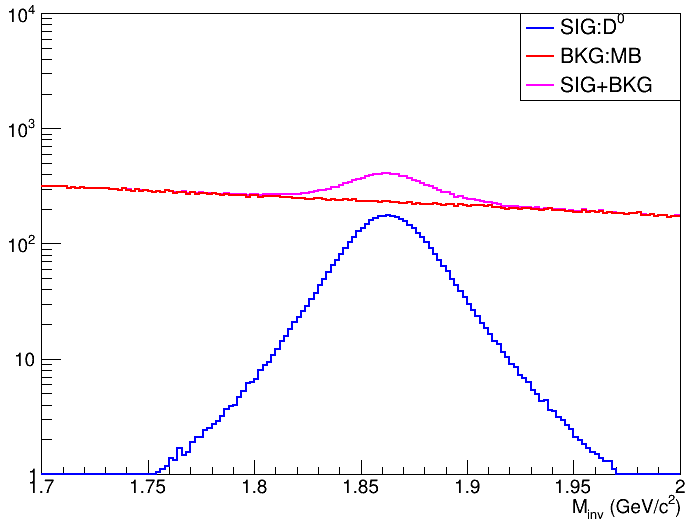}
 \caption{\centering  }
 \label{fig12b}
\end{subfigure}
\caption{(a) Invariant mass spectra of random combinations of pions and kaons and those from $D^0$ decays for $x_F \ge 0.2$. (b) Selections based on the vertex detector used to suppress combinatorial background.}
\label{fig12}
\end{figure}

High precision measurements of the secondary vertex using silicon based central trackers can help reduce the combinatorial random background. Figure (\ref{fig12a}) shows the relative sizes of the background and signal in the pion-kaon invariant mass spectra intended for $D^0$ decay reconstructions for one year of data at the SPD. Figure (\ref{fig12b}) illustrates the effects of the vertex detector in reducing the background. Monte Carlo simulation based studies are in progress to reduce the background further improving the figure of merit making the measurements viable to be compared to theoretical estimates.

\begin{figure}[H]
 \centering
 \includegraphics[width=0.5\textwidth]{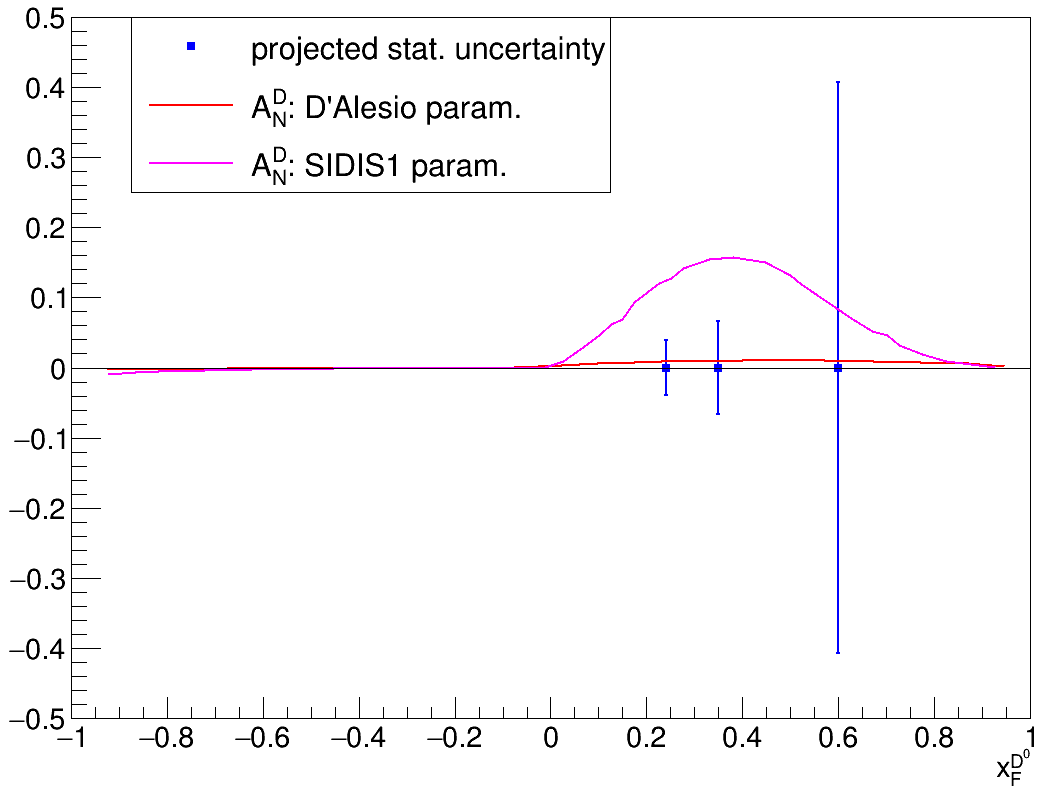}
 \caption{Projected statistical uncertainty of $D^0$ measurements at the SPD for one year of data. \label{fig13c}}
\end{figure}

 As can be observed (Figures \ref{fig13a}, \ref{fig13b}), the two sets of parameters used to describe the Gluon Sivers Function (GSF) in the theoretical calculations predict peak asymmetries differing by an order of magnitude \cite{ref-Saleev_openC}. D'Alesio parameters \cite{ref-GSF-Dalesio} predict asymmetries of the size of $1 \%$ whereas SIDIS1 parameters \cite{ref-GSF-SIDIS1} predict asymmetries of $\sim 17 \%$. SPD measurements can be extremely helpful in reducing such parameter dependence with high enough statistical precision. From the recent Monte Carlo studies of neutral D mesons, the projected statistical uncertainties of the transverse single spin asymmetries for one year of data show (Figure \ref{fig13c}) that measurements at the SPD will provide enough precision to be able to reduce such strong model dependence of theoretical calculations and provide valuable data points for future extraction of the Gluon Sivers Function.

\subsection{Deuteron Measurements}

The SPD will be a unique laboratory to access information about the unpolarized and polarized structure of deuterons as it will have the capacity to collide polarized deuterons over a range of energies. 

\begin{figure}[h]
\begin{subfigure}[h]{0.5\textwidth}
 \includegraphics[width=\textwidth]{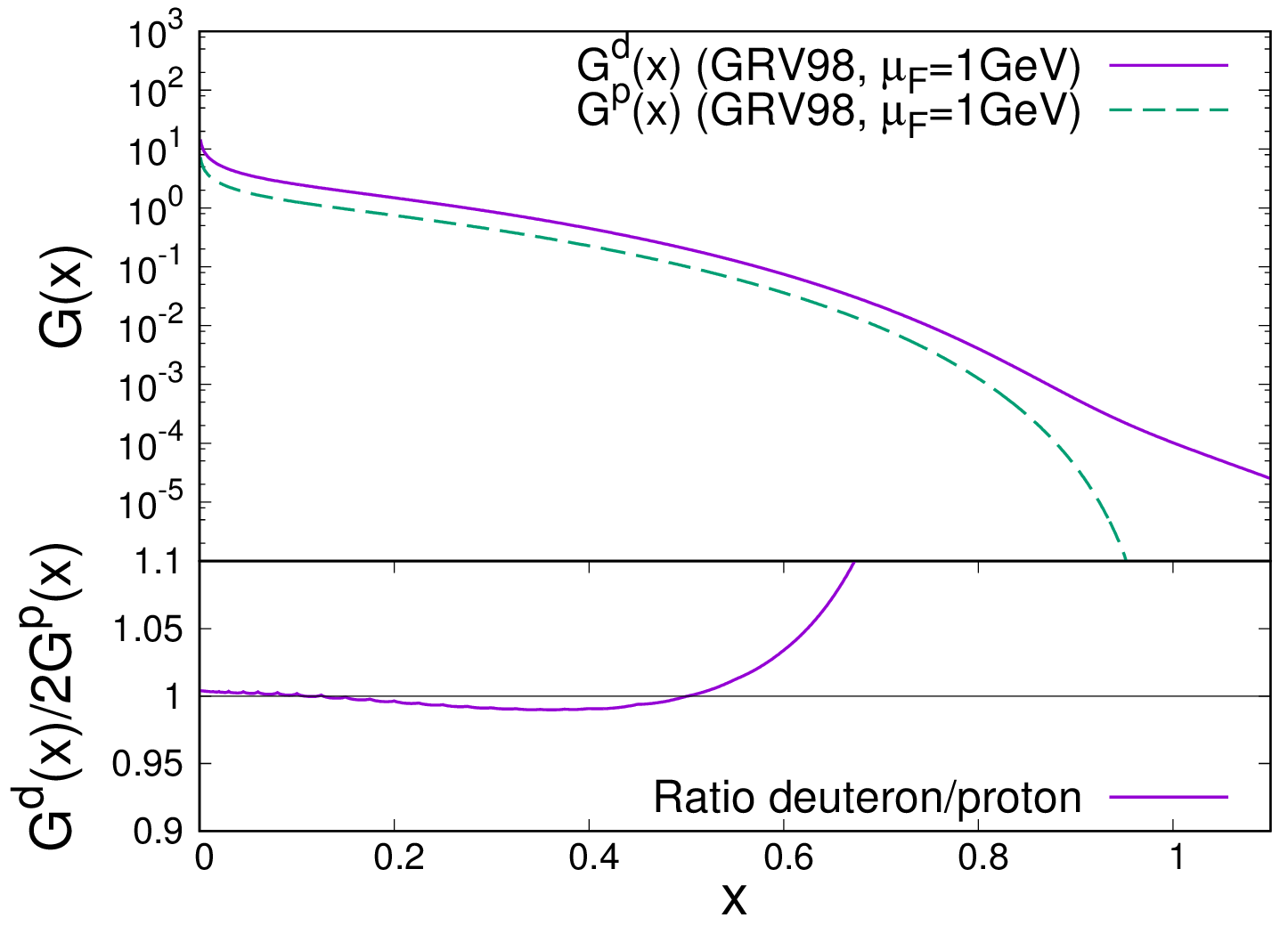}
 \caption{\centering  }
 \label{fig14a}
\end{subfigure}
\hfill
\begin{subfigure}[h]{0.5\textwidth}
 \includegraphics[width=\textwidth]{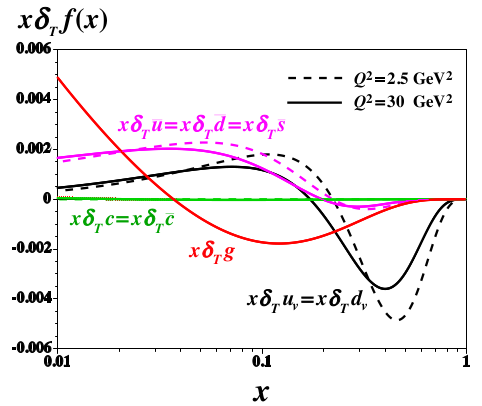}
 \caption{\centering  }
 \label{fig14b}
\end{subfigure}
\caption{(a) Comparisons of the gluon contents of deuterons and protons inside deuterons (\cite{ref-deut-gluon}). (b) Tensor polarized gluon PDF from DGLAP energy evolution of quark/anti-quark PDFs (\cite{ref-stage2}).}
\label{fig14}
\end{figure}

Comparisons of unpolarized gluon PDFs of deuterons and that of protons (Figure \ref{fig14a}) show steep deviations above $x>0.6$ indicating non baryonic contributions. High precision cross-section measurements at the SPD can be compared with theoretical calculations to test the predictions and the size of such deviations.

Tensor polarization of quarks in deuterons has been formerly accessed via asymmetry measurements in DIS experiments (at HERMES). However, Figure \ref{fig14b} shows that DGLAP energy evolution of PDFs suggest that at a higher energy scale (i.e. $Q^2 = 30$ GeV) a non-zero tensor polarized gluon component is possible. Vector and tensor single spin asymmetry measurements at the SPD can test such predictions from the perturbative QCD calculations. 

\section{Discussion}

The SPD experiment at the NICA collider facility at JINR is going to be a unique laboratory to provide a large variety of possible measurements from collisions of polarized proton and deuteron beams over a range of energies and luminosities. In the early stage of operations, measurements at the SPD will probe a wide swathe of interesting physics phenomena encompassing spin effects in low energy nucleon collisions, hyperon and hypernuclei formation, threshold production of charmonia, multi-parton scattering and multi-quark correlations. In the later stage of operations, the SPD experiment will focus on its most prominent goal of accessing the gluon contents inside protons and deuterons via measurements of unpolarized cross-sections and various spin asymmetries in the production of different probe particles. 

Physics programs at the SPD aim to test various phenomena at low to medium energies and provide high precision data to improve present understanding of nucleon structure in general and various spin structures in particular. Results will test QCD in general and will specifically focus on providing data in a kinematic range not probed well as yet to access the gluon content of the nucleons.

Results of detailed Monte Carlo studies are presented in the current work for all three flagship channels of measurements at the SPD aimed at probing gluon content of the nucleons. 

For prompt photons, statistical re-weighting technique of PDFs illustrates the impact of the helicity asymmetry measurements on the $\Delta g(x)$ for one year of data at the SPD. 

For charmonium ($J/\Psi$) also, work presented here illustrates the impact of the measurements of double helicity asymmetries. For both measurements,  results presented here demonstrate that the SPD will have a significant impact in improving our knowledge of the helicity PDF in the large Bjorken-$x$ region as expected from the design and proposal.

For the open-charm channel, work presented in this article show the significant improvement from the early designs by inclusion of the MAPS silicon detector as the central tracking device. Results presented here illustrate the effect of the high precision secondary vertex reconstruction in reducing orders of magnitude higher combinatorial background. The statistical uncertainties of transverse single spin asymmetries have been shown to be able to distinguish between the severe model dependence in the description of the Gluon Sivers Function.

Fixed target Deep Inelastic Scattering experiments have estimated \cite{ref-COMPASS-Siv1, ref-COMPASS-Siv2} separate Sivers and Collins contributions to the transverse single spin asymmetries but $pp$ collider experimental results so far lacked precision to separate between the two effects. For certain probes (i.e. meson production), the SPD will allow investigations of the contributions of Sivers and Collins effects in the single transverse spin asymmetries. A recent analysis \cite{ref-QSivers-Global} of TMD asymmetries measured in various SIDIS  experiments (COMPASS, HERMES) and collider (BRAHMS, STAR) has attempted for the first time to extract Quark Sivers Function. Works \cite{ref-GSivers-Boer, ref-GSivers-NICA} studying the gluon TMD distributions and their contribution to the transverse spin asymmetry measurements of produced hadrons point out the lack of experimental data in this budding field of interest. Attempts to extract gluon Sivers distribution will require data from different kinematic ranges. At present RHIC is the only proton-proton collider capable of colliding polarized beams. In future, the SPD will be able to provide some of the much needed data for such phenomenological global analyses aimed to extract gluon TMD distributions in future.

\section{Conclusion}

The SPD is an international collaboration involving 32 institutes from 14 countries and boasts about three hundred members so far. The collaboration is still growing and is open to participation of experts from different parts of the world.

The conceptual design report (CDR) \cite{ref-CDR} of the experiment was published in early 2021 and was reviewed by the JINR Program Advisory Committee (PAC) in January 2022. Favourable reports from the PAC made it possible for the collaboration to move to the next step of producing a detailed technical design report (TDR) \cite{ref-TDR}.

A tentative schedule expects building of the first stage of the detector to commence in 2026 and possibly take first data sometime around 2028. After a couple of years of data at lower energy and luminosity for the first stage of physics goals, the SPD is scheduled to move to the next stage of upgrades with a focus towards measurements accessing gluon components inside nucleons and light nuclei.

\begin{acknowledgements}
 We would like to thank Alexander Korzenev from the Joint Institute for Nuclear Research for his contributions to the hardware designs of the SPD detector and Alexey Zhemchugov from the Joint Institute for Nuclear Research and Vladimir Andreev from the Lebedev Physical Institute RAS for their valuable contributions in the software infrustructure and simulated data reconstruction.
\end{acknowledgements}


\begin{thebibliography}{9}

\bibitem{ref-EMC}
J.~Ashman et al., ``A measurement of the spin asymmetry and determination of the structure function g1 in deep inelastic muon-proton scatterin'', {\em Phys. Lett. B}, 364-370, 1988.

\bibitem{ref-E704}
D.~L.~Adams \textit{et al.},
``Analyzing power in inclusive pi+ and pi- production at high x(F) with a 200-GeV polarized proton beam'',
{\em Phys. Lett. B} {\bf 264}, 462-466, 1991.

\bibitem{ref-large_AN}
D.~L.~Adams \textit{et al.},
``Comparison of spin asymmetries and cross-sections in pi0 production by 200-GeV polarized anti-protons and protons'',
{\em Phys. Lett. B} \textbf{261}, 201-206, 1991.

\bibitem{ref-QFact}
R.~K.~Ellis et al.,``Factorization and the parton model in QCD'',
{\em Phys. Lett. B} {\bf 78}, 281-284, 1978.

\bibitem{ref-GFact}
J.~C.~Collins et al.,,
``Soft Gluons and Factorization'',
{\em Nucl. Phys. B} \textbf{308}, 833-856, 1988.

\bibitem{ref-NNPDF}
R.~D.~Ball et al.,
``A Determination of parton distributions with faithful uncertainty estimation,''
{\em Nucl. Phys. B} \textbf{809}, 1-63,2009,
[erratum: {\em Nucl. Phys. B} \textbf{816}, 293, 2009]

\bibitem{ref-NNPDF4}
R.~D.~Ball et al.,
``The path to proton structure at 1\% accuracy,''
{\em Eur. Phys. J. C} \textbf{82}, 428, 2022.

\bibitem{ref-DSSV2014}
D.~de Florian et al., ``Evidence for polarization of gluons in the proron'', {\em Phys. Rev. Lett.} {\bf 113}, 012001, 2014.

\bibitem{ref-GRSV}
M.~Gl{\"u}ck et al,
``Models for the polarized parton distributions of the nucleon'',
{\em Phys. Rev. D} {\bf 63}, 094005, 2001. 

\bibitem{ref-BB}
J.~Bl{\"u}mlein et al., 
``Qcd analysis of polarized deep inelastic scattering data'',
{\em Nucl. Phys. B} {\bf 841}, 205-230, 2010.

\bibitem{ref-MSTW}
A.~D.~Martin et al.,
``Parton distributions for the LHC'',
{\em Eur. Phys. J. C} \textbf{63}, 189-285, 2009.

\bibitem{ref-DSSV}
D.~de Florian et al.,
``Global analysis of fragmentation functions for pions and kaons and their uncertainties'',
{\em Phys. Rev. D} {\bf 75}, 114010, 2009.

\bibitem{ref-TMD-overview}
R. Angeles-Martinez et al., ``Transverse momentum dependent (TMD) parton distribution functions: status and overview'' {\em Acta Phys. Polon. B} {\bf 46}, 2501-2534, 2015.

\bibitem{ref-Sivers}
D.~Sivers, ``Single-spin production asymmetries from the hard scattering of pointlike constituents'',
{\em Phys. Rev. D} {\bf 41}, 83-90, 1990.

\bibitem{ref-Collins}
J.~C.~Collins,
``Leading twist single transverse-spin asymmetries: Drell-Yan and deep inelastic scattering,''
{\em Phys. Lett. B} \textbf{536}, 43-48, 2002.

\bibitem{ref-CDR}
V. Abazov et al., ``Conceptual design of the Spin Physics Detector'' \url{https://doi.org/10.48550/arXiv.2102.00442}, 2021.

\bibitem{ref-LHC-AFTER}
C.~Hadjidakis et al.,
``A fixed-target programme at the LHC: Physics case and projected performances for heavy-ion, hadron, spin and astroparticle studies,''
{\em Phys. Rept.} \textbf{911}, 1-83, 2021.

\bibitem{ref-LHCspin}
M.~Santimaria et al.,
``The LHCspin project,''
{\em SciPost Phys. Proc.} \textbf{8}, 050, 2022.

\bibitem{ref-EIC-CDR}
F.~Willeke et al.,``Electron Ion Collider Conceptual Design Report 2021'', doi:10.2172/1765663.

\bibitem{ref-EIC-white}
A.~Accardi et al., ``Electron Ion Collider: The Next QCD Frontier - Understanding the glue that binds us all'', 
\url{https://doi.org/10.48550/arXiv.1212.1701}.

\bibitem{ref-stage1}
A. Abramov et al., {\em Physics of Particles and Nuclei}  {\bf 52}, 1044, 2021.

\bibitem{ref-cHadroprod}
C.~B.~Mariotto et al., ``Soft and hard QCD dynamics in hadroproduction of charmonium'' {\em Eur. Phys. J. C} {\bf 23}, 527-538, 2002.

\bibitem{ref-stage2}
A. Arbuzov et al., {\em Progress in Particle and Nuclear Physics} {\bf 119}, 2021.

\bibitem{ref-DSSVrewt}
D.~De Florian et al.,
``Monte Carlo sampling variant of the DSSV14 set of helicity parton densities'',
{\em Phys. Rev. D} \textbf{100}, 114027, 2019.

\bibitem{ref-GSF-Dalesio}
U.~D'Alesio et al., ``Unraveling the Gluon Sivers Function in Hadronic Collisions at RHIC'', {\em Phys. Rev. D} {\bf 99}, 036013, 2019.

\bibitem{ref-GSF-SIDIS1}
U.~D'Alesio et al., ``Unraveling the Gluon Sivers Function in Hadronic Collisions at RHIC'', {\em JHEP} {\bf 119}, 1509, 2015.

\bibitem{ref-DSSVpvt}
W. Vogelsang (Universitat Tubingen), R. Sassot (Universidad de Buenos Aires - Exactas), I. Borsa. (Universidad de Buenos Aires - Exactas), Personal Communication, 2021.

\bibitem{ref-hypernuc}
Jean-Marc Richard et al., ``Lightest neutral hypernuclei with strangeness −1 and −2'' {\em Phys. Rev. C} {\bf 91}, 014003, 2014.

\bibitem{ref-TDR}
V. Abazov et al., Technical design of the Spin Physics Detector. {\em To be published}.

\bibitem{ref-jpsi-AN}
A. Karpishkov et al., ``Estimates for the single-spin asymmetries in the $p^{\uparrow}p \rightarrow J/\Psi X$ process at PHENIX RHIC and SPD NICA'' {\em Phys. Rev. D} {\bf 104}, 016008, 2021.

\bibitem{ref-COMPASS-openC}
C.~Adolph et al.,
``Leading and Next-to-Leading Order Gluon Polarization in the Nucleon and Longitudinal Double Spin Asymmetries from Open Charm Muoproduction,''
{\em Phys. Rev. D} \textbf{87}, 052018, 2013.

\bibitem{ref-Saleev_openC}
A.~Karpishkov and V.~Saleev,
``On transverse single-spin asymmetries in $D$-meson production at the SPD NICA experiment,''
\url{arXiv:2212.07636}.

\bibitem{ref-deut-gluon}
S. Broadsky et al., ``The gluon and charm content of the deuteron'' {\em Phys. Lett. B} {\bf 783}, 287-293, 2018.

\bibitem{ref-COMPASS-Siv1}
M.~G.~Alekseev \textit{et al.},
``Measurement of the Collins and Sivers asymmetries on transversely polarised protons'',
{\em Phys. Lett. B} {\bf 692}, 240-246, 2010.

\bibitem{ref-COMPASS-Siv2}
M.~G.~Alekseev \textit{et al.},
``Measurement of $P_T$-weighted Sivers asymmetries in leptoproduction of hadrons'',
{\em Nucl. Phys. B} {\bf 940}, 34-53, 2019.

\bibitem{ref-QSivers-Global}
J.~Cammarota et al., ``Origin of single-spin asymmetries in high-energy collisions'', {\em Phys. Rev. D} {\bf 102}, 054002, 2020.

\bibitem{ref-GSivers-NICA}
U.~D'Alesio et al., ``Investigating the Transverse Momentum Dependent Gluon Sivers Function in Quarkonium Production at pp Colliders'', {\em Few-Body Systems.} {\bf 62}, 2021.

\bibitem{ref-GSivers-Boer}
D.~Boer et al., ``The Gluon Sivers Distribution:Status and Future Prospects'', \url{https://doi.org/10.1155/2015/371396}

\end{thebibliography}
\end{document}